\pdfoutput=1

\documentclass[twocolumn,showpacs,amssymb,aps,nofootinbib,floatfix,superscriptaddress]{revtex4-1}

\usepackage{epsfig}
\usepackage{hyperref}

\begin{document} 

\title{\bf Wounded quarks in A+A, p+A, and p+p collisions}

\author{Piotr Bo\.zek}
\email{Piotr.Bozek@fis.agh.edu.pl}
\affiliation{AGH University of Science and Technology, Faculty of Physics and
Applied Computer Science, al. Mickiewicza 30, 30-059 Cracow, Poland}

\author{Wojciech Broniowski}
\email{Wojciech.Broniowski@ifj.edu.pl}
\affiliation{The H. Niewodnicza\'nski Institute of Nuclear Physics, Polish Academy of Sciences, 31-342 Cracow, Poland}
\affiliation{Institute of Physics, Jan Kochanowski University, 25-406 Kielce, Poland}

\author{Maciej Rybczy\'nski}
\email{Maciej.Rybczynski@ujk.edu.pl}
\affiliation{Institute of Physics, Jan Kochanowski University, 25-406 Kielce, Poland}

\begin{abstract}
We explore predictions of the wounded quark model for particle production and properties of the initial state formed
in ultrarelativistic heavy-ion collisions. The approach is applied uniformly to A+A collisions in a wide collision 
energy range, as well as for p+A and p+p collisions at the CERN Large Hadron Collider (LHC). We find that generically the 
predictions from wounded quarks for such features as eccentricities or initial sizes are close (within 15\%) to predictions of the 
wounded nucleon model with an amended 
binary component. A larger difference is found for the size in p+Pb system, where the wounded quark model yields a smaller (more compact)
initial fireball than the standard wounded nucleon model. The inclusion of subnucleonic degrees of freedom allows us to analyze p+p collisions
in an analogous way, with predictions that can be used in further collective evolution. The approximate linear dependence of particle production in A+A collisions 
on the number of wounded quarks, as found in previous studies, makes the approach based on wounded quarks natural. Importantly, at the LHC energies 
we find approximate uniformity in particle production 
from wounded quarks, where at a given collision energy per nucleon pair similar production of initial entropy per source 
is needed to explain the particle production from p+p collisions up to A+A collisions. We also discuss the sensitivity of the wounded quark model predictions 
to distribution of quarks in nucleons, distribution of nucleons in nuclei, and to the quark-quark inelasticity profile in the impact parameter. 
In our procedure, the quark-quark inelasticity profile is chosen in such a way that the experiment-based parametrization of the proton-proton 
inelasticity profile is properly reproduced. The parameters of the overlaid multiplicity distribution is fixed from p+p and p+Pb data.
\end{abstract}

\date{ver. 2, 3 May 2016}



\maketitle

\section{Introduction}

The idea of wounded quarks~\cite{Bialas:1977en,*Bialas:1977xp,*Bialas:1978ze,Anisovich:1977av} was proposed shortly after the 
successful concept of wounded nucleons~\cite{Bialas:1976ed}, in the quest 
of understanding in a natural way particle production in high-energy nuclear collisions. 
In particular, in these early applications one looked for appropriate scaling of particle production with the number of participants, and the idea of {\em wounding}, i.e., production from a participant
independent on the number of its collision with participants from the other nucleus, proved very useful. It is based on the Glauber approach~\cite{Glauber:1959aa} adapted to inelastic 
collisions~\cite{Czyz:1969jg}, and the soft particle production may be justified in terms of the Landau-Pomeranchuk effect (for review see~\cite{Bialas:2008zza}).

With the advent of the BNL Relativistic Heavy-Ion Collider (RHIC),
 the idea of wounded nucleons was revived by the PHOBOS Collaboration~\cite{Back:2001xy}. However, to explain the multiplicity distribution in Au+Au collisions 
it was necessary to include a component proportional to the binary collisions~\cite{Kharzeev:2000ph}, i.e., depending nonlinearly on the number of the wounded nucleons. 
Then the average number
of particles produced at mid-rapidity is
\begin{eqnarray}
\frac{dN_{\rm ch}}{d\eta} \sim \frac{1-\alpha}{2}N_{\rm W} + \alpha N_{\rm bin}, \label{eq:wn}
\end{eqnarray}
where $N_{\rm W}$ denotes the number of wounded nucleons and  $N_{\rm bin}$ the number of binary collisions, with the first 
term interpreted as soft, and the second as hard production~\cite{Kharzeev:2000ph}. 

In heavy-ion phenomenology, the entropy density profile in the initial state is the most important source of uncertainty in hydrodynamic models. Besides the 
Monte Carlo Glauber initial condition, IP-Glasma and KLN initial conditions are frequently
used~\cite{Schenke:2012wb,Drescher:2006pi}. One should note that differences in the parameterizations of the initial entropy deposition give different  
eccentricities of the fireball~\cite{Moreland:2014oya}. Within the Monte Carlo Glauber approach, an understanding
in more microscopic terms of the phenomenological binary collisions component in Eq.~(\ref{eq:wn}) is desired, 
whereby uncertainties in the initial entropy deposition in the Glauber model should be reduced.

Eremin and Voloshin~\cite{Eremin:2003qn} noticed that the RHIC multiplicity data can be naturally reproduced within a wounded quark model. With a larger number of constituents (three with quarks) 
and a  reduced cross section of quarks compared to nucleons one can obtain an approximately linear increase in particle 
production with the number of wounded quarks, denoted as $Q_{\rm W}$ in this paper, namely:
\begin{eqnarray}
\frac{dN_{\rm ch}}{d\eta} \sim Q_{\rm W}. \label{eq:wq}
\end{eqnarray}
The quark scaling has also been reported for the SPS energies~\cite{KumarNetrakanti:2004ym}.
It has  been vigorously promoted the PHENIX Collaboration~\cite{Adler:2013aqf,Adare:2015bua}. Experimental data on particle production has been compared to 
the number of wounded nucleons or wounded quarks. The number of wounded quarks or nucleons in a collisions is obtained from  a Glauber Monte Carlo code \cite{Alver:2008aq,Loizides:2014vua}, where for the model with quark constituents additionally three quarks are distributed in each nucleon and collisions occur between pairs of quarks.  An approximate,  
uniform scaling of  the multiplicity and of the 
transverse energy  with the number of wounded quarks is claimed~\cite{Adler:2013aqf,Adare:2015bua}, also including the LHC data~\cite{Lacey:2016hqy}.

The basic effect of the subnucleonic degrees of freedom in particle production is a stronger combinatorics, which accomplishes 
the approximately linear scaling of production with the number of constituents.
An intermediate combinatorics is realized in the quark-diquark model~\cite{Bialas:2006kw,*Bialas:2007eg}, which led to proper description of the RHIC data as well as the proton-proton 
scattering amplitude (including the differential elastic cross section) at the CERN Intersecting Storage Rings (ISR) energies. 

The purpose of this paper is to explore in detail predictions of the wounded quark approach to particle production and properties of the initial state, as well as to 
investigate sensitivity to the assumptions concerning the distribution of quarks and the quark-quark collision profile. While this work was nearing completion, 
a similar study for A+A collisions by Zheng and Yin~\cite{Zheng:2016nxx} appeared. 
 In Ref.~\cite{Lacey:2016hqy}
the scaling of particle production with the number of quarks is discussed, whereas in Ref.~\cite{Mitchell:2016jio} different ways of generating quark positions in the nucleon are compared. In addition, Loizides~\cite{Loizides:2016djv} presented a detailed study of the initial state in heavy-ion collisions for models
involving a number of partons in a nucleon different than three.

The difference in this work compared to other studies is that our methodology fixes the proton shape and the quark-quark 
inelasticity profile to reproduce accurately the inelasticity profile
in proton-proton collisions. It is important, since the details of the model of partons in nuclei, such as  the effective size of the nucleon, the quark-quark
cross section, and the shape of the inelasticity profile in quark-quark scattering turn out to be important for the results.
With the fitted parameters of the quark distribution in nucleons and quark scattering, we calculate the basic characteristics 
(eccentricities, sizes, and their event-by-event fluctuations) of the initial state in A+A 
collisions. We also apply the approach to p+A and p+p collisions, not analyzed in other works. This extension is important, 
as it checks consistency of the constituent quark scaling, as well as probes the limit of collectivity in small systems.

The outline of our paper is as follows: In Sec.~\ref{sec:qconst} we present our method of fixing the parameters of the model in terms of the 
proton-proton scattering amplitude, with further details given in Appendix~\ref{sec:inel}. Next, we discuss the overlaying of fluctuations on the Glauber sources, with 
the gamma distribution in the initial stage and the Poisson distribution at hadronization, convolving to the popularly used negative binomial distribution. These extra fluctuations are necessary 
to describe the p+A and p+p collisions, or the A+A collisions at lowest centralities; they also contribute to eccentricities or event-by-event fluctuation measures. 
In Sec.~\ref{sec:AA} we look at the A+A collisions at a wide energy range, testing the wounded quark scaling of hadron production and its sensitivity to model assumptions.
We then pass to the study of eccentricities (including the hypercentral U+U collisions) and size fluctuations. Section~\ref{sec:pA} presents the results for p+A collisions, where we 
use the measured multiplicity distributions of produced hadrons to fix the parameters of the overlaid distribution. We then obtain the eccentricities and sizes of the initial fireball. We find that in the case of 
wounded quarks the initial size is significantly more compact than in the standard wounded nucleon model. In Sec.~\ref{sec:npp} we analyze in an 
analogous way the p+p collisions, which is possible with subnucleonic degrees of freedom. The obtained initial fireballs for highest-multiplicity p+p collisions 
could be used in studies of subsequent collective evolution of the system.
The possibility of including more wounded constituents is explored in Sec.~\ref{sec:more}.
Finally, Sec.~\ref{sec:concl} contains our summary, with the main conclusion that the concept of wounded quarks works 
uniformly from p+p to Pb+Pb collisions at the LHC energies, providing approximately linear scaling of hadron production with the number of sources, and resulting in 
properties of the initial fireball close to the predictions with wounded nucleons amended with binary collisions. At RHIC, we observe dependence of the scaling on the colliding nuclei, and 
substantial difference with the p+p channel (about 30\%), which indicates the need for improvement of the theoretical description at lower collision energies.

\section{General formulation}

\subsection{Quarks inside nucleons \label{sec:qconst}}

Constituent quarks are localized within nucleons, which results in their clustering as opposed to free, unconfined distribution in the nucleus. 
For that reason we attempt to model their distributions in a realistic way. Firstly, we randomly place the centers of quarks in the nucleons according to the 
radial distribution 
\begin{eqnarray}
\rho(r)=\frac{r^2}{r_0^3} e^{-r/r_0} \ ,  \label{eq:eps}
\end{eqnarray}
with a shift to the center of mass of the nucleon after generating the positions of the three quarks.
Positions of nucleons in a nucleus are distributed according to a standard Woods-Saxon radial density profile with 
appropriate parameters~\cite{Broniowski:2007nz,Shou:2014eya}, or are taken from outside microscopic calculations, such as~\cite{Alvioli:2009ab}.
Nuclear deformation~\cite{Heinz:2004ir,Filip:2007tj,Filip:2009zz,Rybczynski:2012av} is incorporated for deformed nuclei, such as $^{63}$Cu, $^{197}$Au, or $^{238}$U, used at RHIC.

The quark-quark wounding profile (inelasticity profile in the impact parameter $b$) is for simplicity assumed to have the Gaussian shape
\begin{eqnarray}
p^{\rm in}_{\rm qq}(b)=2\pi b e^{-\pi b^2/\sigma_{qq}},  \label{eq:gauss}
\end{eqnarray}
where $\sigma_{qq}$ is the quark-quark inelastic cross section. 

The choice of this profile, together with the parametrization~(\ref{eq:eps}), is important, as it determines the probability of a collision at the transverse separation $b$, which 
influences the wounded quark scaling and the properties of the formed initial state.
The numbers $r_0$ and  $\sigma_{qq}$ are treated as adjustable parameters and, at each collision energy, are fitted to the nucleon-nucleon total inelastic cross section and to the nucleon-nucleon 
inelasticity profile. The latter can be obtained straightforwardly from working parameterizations of the NN scattering data, which describe both inelastic and elastic collision amplitudes. 
Here we take the parametrization of Ref.~\cite{Fagundes:2013aja} based on the 
Barger-Phillips model~\cite{Phillips:1974vt}. The details of our procedure are given in Appendix~\ref{sec:inel}. The bottom line of this procedure is that we reproduce with sufficient accuracy the 
experimental inelasticity profile of the NN collisions. The values of our parameters are listed in Table~\ref{tab:inel} in Appendix~\ref{sec:inel}.

In the wounded quark Glauber model of the  PHENIX Collaboration~\cite{Adler:2013aqf,Adare:2015bua}, 
the size of the proton is energy independent, whereas the quark-quark cross section increases with the collision energy. In a different 
approach, the increase of the nucleon-nucleon cross section with energy is achieved with the increase 
of the nucleon size~\cite{Heinz:2011mh}. When fitting the inelastic profile 
at different energies in our approach we find that the  effective size of the nucleon increases very weakly with the energy, 
while the quark-quark cross section increases sizeably with the energy. 
The corresponding change in the quark-quark inelasticity profile is determined by $\sigma_{qq}$, cf. Eq.~(\ref{eq:gauss}).

\subsection{Overlaid distribution \label{sec:over}}

The number of particles produced in p+p collisions fluctuates. In the wounded quark model only a part of these fluctuations
can be accounted for by fluctuations of the number of wounded quarks in a collisions. To describe the experimentally measured distributions, 
the charged hadron distribution should be written as a convolution of the numbers of hadrons $n_k$ produced from each wounded quark,
\begin{eqnarray}
&& P(n)= \sum_{i} P_{wq}(i) \times \\ && \sum_{n_1, n_2, \dots, n_i} P_{hq}(n_1) P_{hq}(n_2)  \dots P_{hq}(n_i) \delta_{n,n_1+n_2+\dots+n_i}, \nonumber
\end{eqnarray}
where $P_{wq}$ is the distribution of the number $i$ of wounded quarks and $P_{hq}$ is the overlaid 
distribution of the number of hadrons from an individual wounded quark. 
The above formula should be used uniformly for p+p, p+A, and A+A collisions if the production 
mechanism is to be universal. In the following we use the negative binomial distribution to parametrize $P_{hq}$, namely
\begin{equation}
P_{NB}(n|\bar{n},\kappa)=\frac{\Gamma(n+\kappa){\bar{n}}^n\kappa^\kappa}{\Gamma(\kappa)n! (\bar{n}+\kappa)^{n+\kappa}}\ ,
\label{eq:NB}
\end{equation}
where  $\Gamma(z)$ is the Euler Gamma function, the average is given by $\bar{n}$, and $\kappa=\bar{n}^2/({\rm var}(n)-\bar{n})$
(larger $\kappa$ means smaller fluctuations).
This form has been widely used to describe the multiplicity distributions in p+p collisions~\cite{Giovannini:1985mz}.
The parameters $\bar{n}$ and $\kappa$ of the negative binomial distribution can be adjusted to reproduce 
the observed multiplicity distribution in p+p (Sec.~\ref{sec:npp}) or  p+A  collisions (Sec.~\ref{sec:mult_pA}).
In collisions of heavy nuclei the multiplicity distribution, or the mean multiplicity in each centrality bin, 
is almost independent on the overlaid distribution, except for very central collisions where it should be taken into account.
However, effects of the overlaid distribution show up  also in  A+A collisions in various event-by-event fluctuation quantities.

When dividing the dynamics of the collision into two stages, namely 1)~the formation of the initial fireball
and 2)~its subsequent evolution and decay into hadrons, one should also separate the multiplicity fluctuations into two corresponding parts.
The first part comes from the fluctuations in the initial entropy deposition in the fireball, whereas
the other part comes from entropy  production during the expansion of the fireball and subsequent hadronization. 
Note that a smooth, linear increase of  entropy 
in viscous hydrodynamics is effectively included in the normalization 
coefficient. We neglect possible fluctuations in the hydrodynamic phase~\cite{Kapusta:2011gt,Gavin:2008ev} (which are expected not to be 
very large~\cite{Yan:2015lfa}) and from correlated  particle emissions at freeze-out. 
We follow the usual  assumption that the particle
emission is given by a Poisson process, with the mean proportional to the entropy in the fluid element.

As the negative binomial distribution is a convolution of the gamma and the Poisson distributions, 
the entropy distribution $P(s)$ in the fireball is given by a convolution of a gamma distribution,
\begin{equation}
P_\Gamma(s|\bar{n},\kappa)=\frac{s^{\kappa-1}\kappa^\kappa}{\Gamma(\kappa)\bar{n}^\kappa}e^{-\kappa s/\bar{n}}, \label{eq:Gamma}
\end{equation}
with the distribution of the number of wounded quarks,
\begin{equation}
P(s)= \sum_i P_{wq}(i) P_{\Gamma}(s|i \bar{n}, i\kappa).
\end{equation}
Microscopically, it means that each wounded quark deposits a random entropy $s$ taken from the gamma distribution
(\ref{eq:Gamma}), with same parameters as the parameters of the negative binomial distribution adjusted to the multiplicity distributions. 
The fluctuations in entropy deposition increase the fluctuations and deformations of the initial fireball. The effect is especially important in small systems or for
central A+A collisions \cite{Rybczynski:2012av,Rybczynski:2013yba,Bozek:2013uha,Kozlov:2014fqa}.
Throughout this work, according to the above discussion, we overlay a negative binomial distribution on top of the wounded quark distribution when 
calculating the hadron multiplicity distribution in p+p and p+A collisions, 
and we overlay a gamma distribution when calculating the entropy profile of the initial fireball and its properties, such as eccentricities or size.

\subsection{Smearing of sources}

On physical grounds, the Glauber sources must possess a certain transverse size; smoothness of the initial condition is also required by hydrodynamics. For that reason 
one needs to smooth the initial distribution. We follow the Gaussian prescription, where the entropy density from a single source centered at a transverse point $(x_0,y_0)$ 
is given with the profile 
\begin{eqnarray}
g(x,y)=\frac{1}{2\pi \sigma^2} \exp \left ( - \frac{(x-x_0)^2+(y-y_0)^2}{2\sigma^2}\right ).  \label{eq:smooth}
\end{eqnarray}
Unless otherwise stated, we use arbitrarily $\sigma=0.2$~fm in the wounded quark model and $\sigma=0.4$~fm in the wounded nucleon model.
One should note that in small systems the fireball eccentricities or sizes depend sensitively on the value of $\sigma$.

\section{A+A collisions \label{sec:AA}}

\subsection{Multiplicity distribution}

We begin the presentation of our results 
by showing the outcome of the wounded quark model for the A+A collisions. The results, obviously, depend on the centrality of the collision, which is defined experimentally 
via the response of detectors, hence its detailed modeling may be complicated. Assuming that the centrality is obtained from the multiplicity of the collision, we may 
evaluate it as percentiles of the event-by-event distribution of the number of sources, which is accurate enough except for the very central and very peripheral events~\cite{Broniowski:2001ei}. 
A better way is to obtain centrality from the number of sources with weights from an overlaid distribution (see Sec.~\ref{sec:over}). However, in presenting the results in the field it is customary to use the 
number of wounded {\em nucleons} $N_{\rm W}$, as obtained from the Glauber simulations with the model of Eq.~(\ref{eq:wn}). 
We follow this convention, thus $N_{\rm W}$ serves as a label the centrality classes, even when these are determined by the wounded quarks.

\begin{figure}
\begin{center}
\includegraphics[width=0.44 \textwidth]{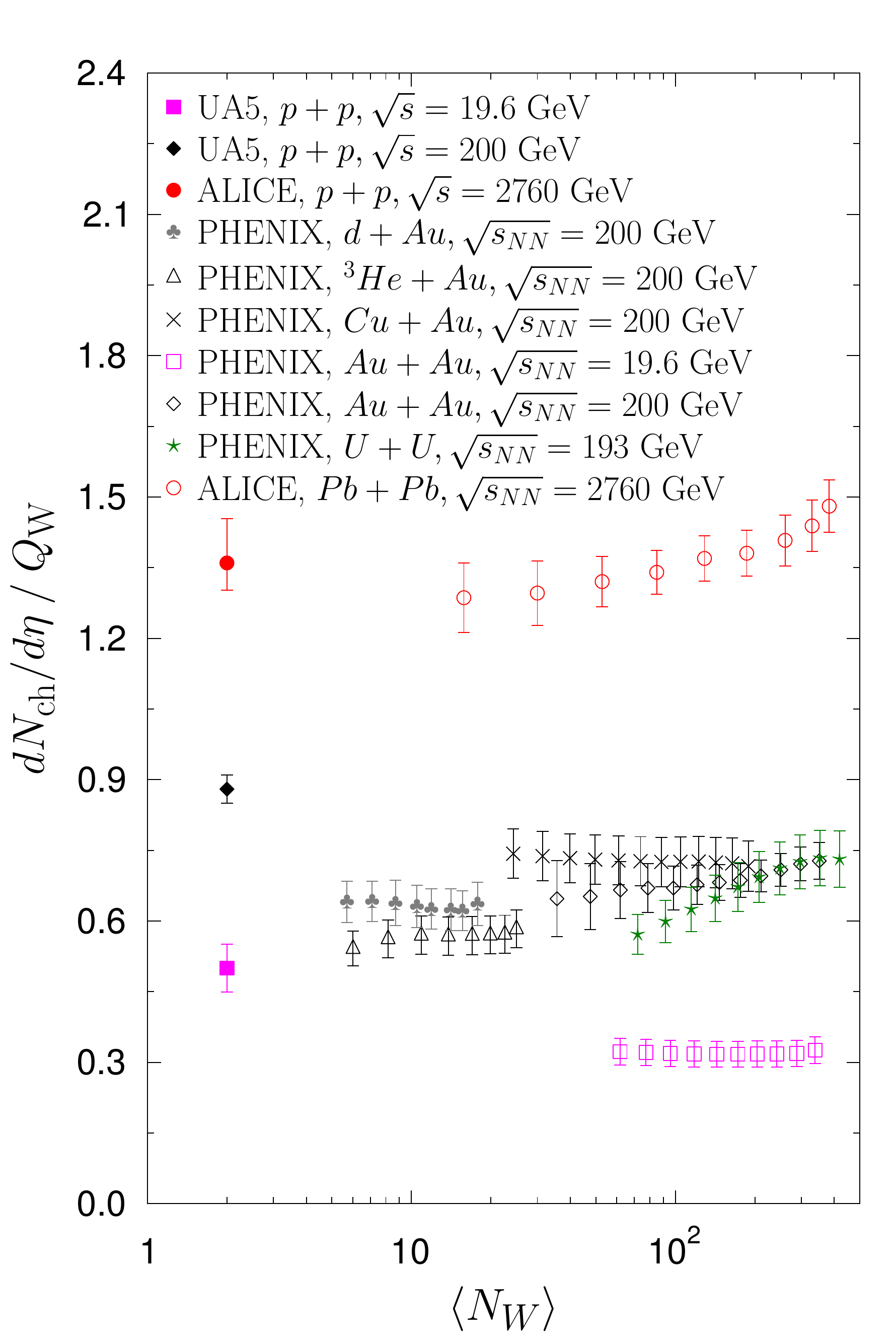} 
\end{center}
\vspace{-5mm}
\caption{Experimental multiplicity of charged hadrons per unit of pseudorapidity (at mid-rapidity) divided by the number of wounded quarks, $dN_{\rm ch}/d\eta/Q_{\rm W}$, 
plotted as function of centrality expressed via the number of wounded 
nucleons.  We also show the results for the p+p collisions (filled symbols at $\langle N_{\rm W} \rangle=2$) discussed in Sec.~\ref{sec:npp}.  The data are from
Refs.~\cite{Alner:1986xu,Adam:2015gka,Aamodt:2010cz,Adare:2015bua}. \label{fig:dndeta_scaled}}
\end{figure}

The basic test of the wounded quark model is its very definition of Eq.~(\ref{eq:wq}), whereby the production of hadrons should be proportional to the number of wounded quarks $Q_{\rm W}$.
In particular, the ratio of  the multiplicity of charged hadrons at mid-rapidity $dN_{\rm ch}/d\eta$ divided by the number of wounded $Q_{\rm W}$, 
should not depend on centrality or on colliding nuclei. To find 
$Q_{\rm W}$ corresponding to a given centrality we have carried out simulations in the Glauber model with the help of appropriately modified GLISSANDO code~\cite{Broniowski:2007nz,Rybczynski:2013yba}. 
In the standard calculation, the nuclear profiles of heavy nuclei are obtained from the Woods-Saxon form with the 
nucleon-nucleon expulsion radius of 0.9~fm~\cite{Broniowski:2010jd}. For the deuteron we use the Hulthen wave function, and for $^3$He we take the distributions 
from the method of Ref.~\cite{Carlson:1997qn} as provided in Ref.~\cite{Loizides:2014vua}. The quarks are then generated in nucleons according to Eq.~(\ref{eq:eps}), whereas the 
inelasticity profile for the quark-quark collisions is of the Gaussian form~(\ref{eq:gauss}). 

The results are shown in Fig.~\ref{fig:dndeta_scaled}. We note that, in qualitative agreement with the earlier studies~\cite{Eremin:2003qn,KumarNetrakanti:2004ym,Adler:2013aqf,Adare:2015bua}, 
the dependence of $dN_{\rm ch}/d\eta/Q_{\rm W}$ on centrality is approximately flat, and increases with the collision energy.  
We note that some deviation from a linear scaling between the initial entropy and the final particle density in pseudorapidity is possible due to different mean transverse momentum or 
different entropy production at different centralities~\cite{Song:2008si} during the evolution of the system.
In contrast, the flatness is not the case of the wounded nucleon model of Eq.~(\ref{eq:wn}), where, as is well known, 
the ratio $dN_{\rm ch}/d\eta/N_{\rm W}$ increases considerably with $N_{\rm w}$.
Moreover, the value of  $dN_{\rm ch}/d\eta/Q_{\rm W}$ (i.e., the average number of charge hadrons per unit of rapidity coming from a single wounded 
quark), is at a given energy roughly similar for various considered reactions. 
For the LHC energies it is also consistent with the p+p collisions, where the data are taken from Ref.~\cite{Adam:2015gka} and $Q_{\rm W}$ is 
obtained in Sec.~\ref{sec:npp}. 
At RHIC energies of $\sqrt{s_{NN}}=200$ and $19.6$~GeV the p+p point, obtained with the data from Ref.~\cite{Alner:1986xu}, is noticeably higher (about 30\%) than the band for the A+A collisions.
Also, the dependence on the colliding nuclei at RHIC indicates a systematic uncertainly of the approach.

The above discussed universality of the hadron production in  p+p vs A+A,  holding fairly well for the LHC and to a lesser degree for the lower energies, is an important check of the idea of the 
independent production from the wounded quark sources. We note that at lower energies it is not very accurate. Also, as the plots in Fig.~\ref{fig:dndeta_scaled} are not really flat, 
the wounded quark scaling is approximate.  This observation may suggest that the effective number of subnucleonic degrees of freedom at RHIC energies is smaller that three, as implied, e.g., by the 
quark-diquark model  \cite{Bialas:2006kw}, and at higher energies may be larger than three (cf. Sec.~\ref{sec:more}). 

\begin{figure}
\begin{center}
\includegraphics[width=0.44 \textwidth]{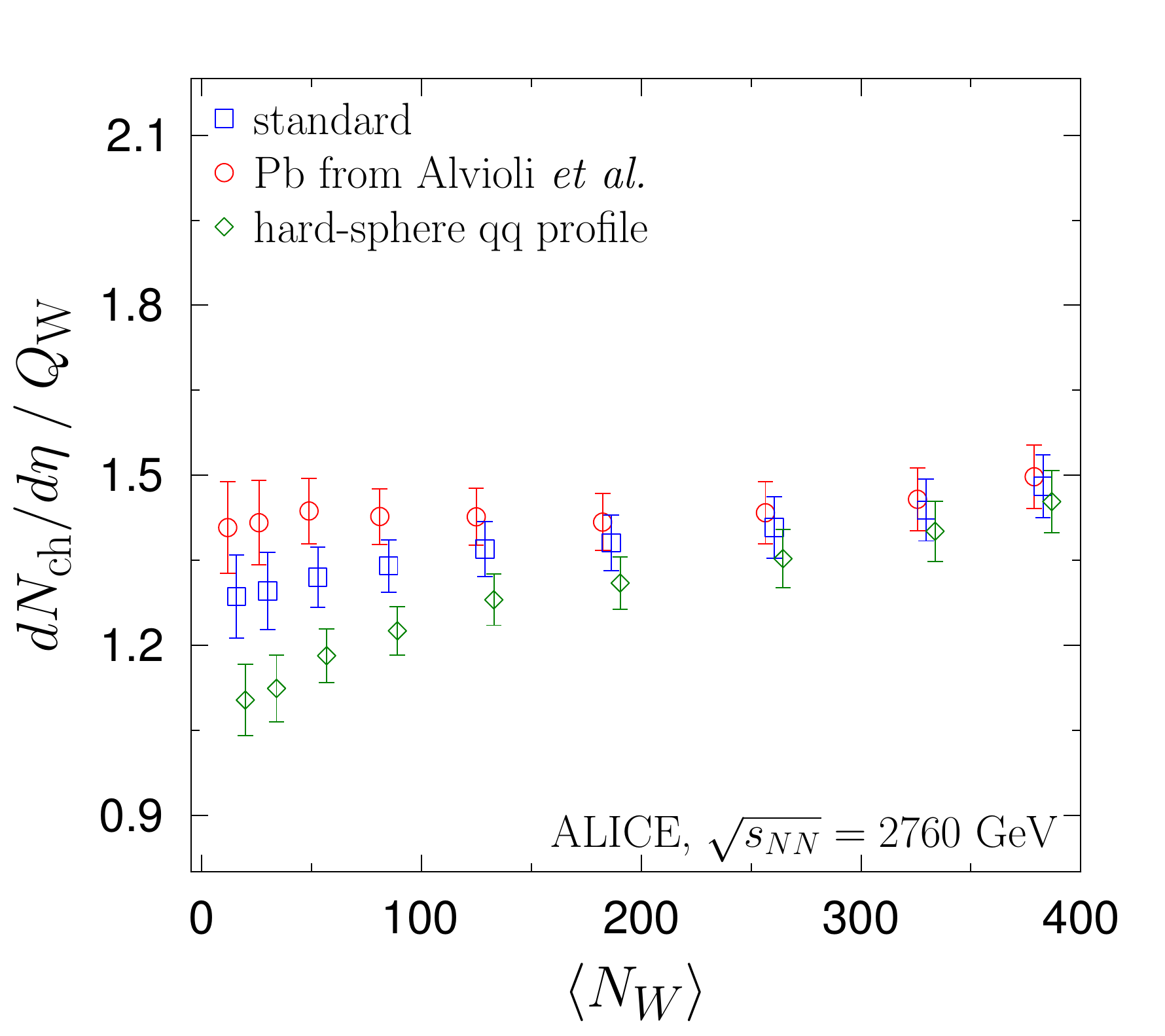} 
\end{center}
\vspace{-5mm}
\caption{Comparison of $dN_{\rm ch}/d\eta/Q_{\rm W}$ in Pb+Pb collisions at $\sqrt{s_{NN}}=2.76$~TeV \cite{Aamodt:2010cz}, obtained with the standard GLISSANDO simulations
with the nuclear Woods-Saxon distribution (squares), compared to the case of 
correlated nuclear distributions 
of Alvioli et al. from  Ref.~\cite{Alvioli:2009ab}.   We also show the calculation with quarks implemented as in the 
Glauber Monte Carlo of Refs.~\cite{Adare:2015bua,Loizides:2016djv} (diamonds), i.e., with the nucleon profile from Ref.~\cite{Adare:2015bua}
 and  a  hard-sphere 
quark-quark inelasticity profile. \label{fig:gauss_vs_hs}}
\end{figure}

To assess the sensitivity of  the approximate 
wounded quark scaling for the particle production, we check the dependence on the nucleon distributions in nuclei, as well as on the 
treatment of quarks. First, in Fig.~\ref{fig:gauss_vs_hs} we compare our standard GLISSANDO calculation described in Sec.~\ref{sec:qconst} 
with Woods-Saxon distributions (squares) to an analogous simulation with
distributions obtained from Ref.~\cite{Alvioli:2009ab}, where central two-body nucleon-nucleon correlations are incorporated (circles). We note that the latter case yields even 
more flat result as a function of centrality.
The reason for this behavior is a longer tail in the one-body nucleon distributions of Ref.~\cite{Alvioli:2009ab} as compared to the Woods-Saxon profile~\cite{Broniowski:2007nz}.

Second, we compare our standard results to the calculation made as in the quark Glauber Monte Carlo code described in Ref.~\cite{Adare:2015bua,Loizides:2016djv} (diamonds), 
i.e., with a modified distribution of quarks in the nucleon and  a hard-sphere 
quark-quark inelasticity profile (the resulting nucleon-nucleon inelasticity profile in this case is different from the experimental parametrization used in this paper, cf. Appendix \ref{sec:inel}).
We notice a substantial difference from our standard result, with a larger breaking of the wounded quark scaling. 
The conclusion emerging from Fig.~\ref{fig:gauss_vs_hs} is that the modeling 
of the subnucleonic structure should be done as accurately and realistically as possible. The same concerns the nuclear distributions, the distributions of 
quarks inside nucleons, or the quark-quark inelasticity profiles, as such details influence the bulk properties in A+A collisions. Still  
some systematic errors, stemming from the model assumptions on the nuclear and subnucleonic structure of the nuclei, are unavoidable. 

\subsection{Eccentricities}

\begin{figure}
\begin{center}
\includegraphics[width=0.45 \textwidth]{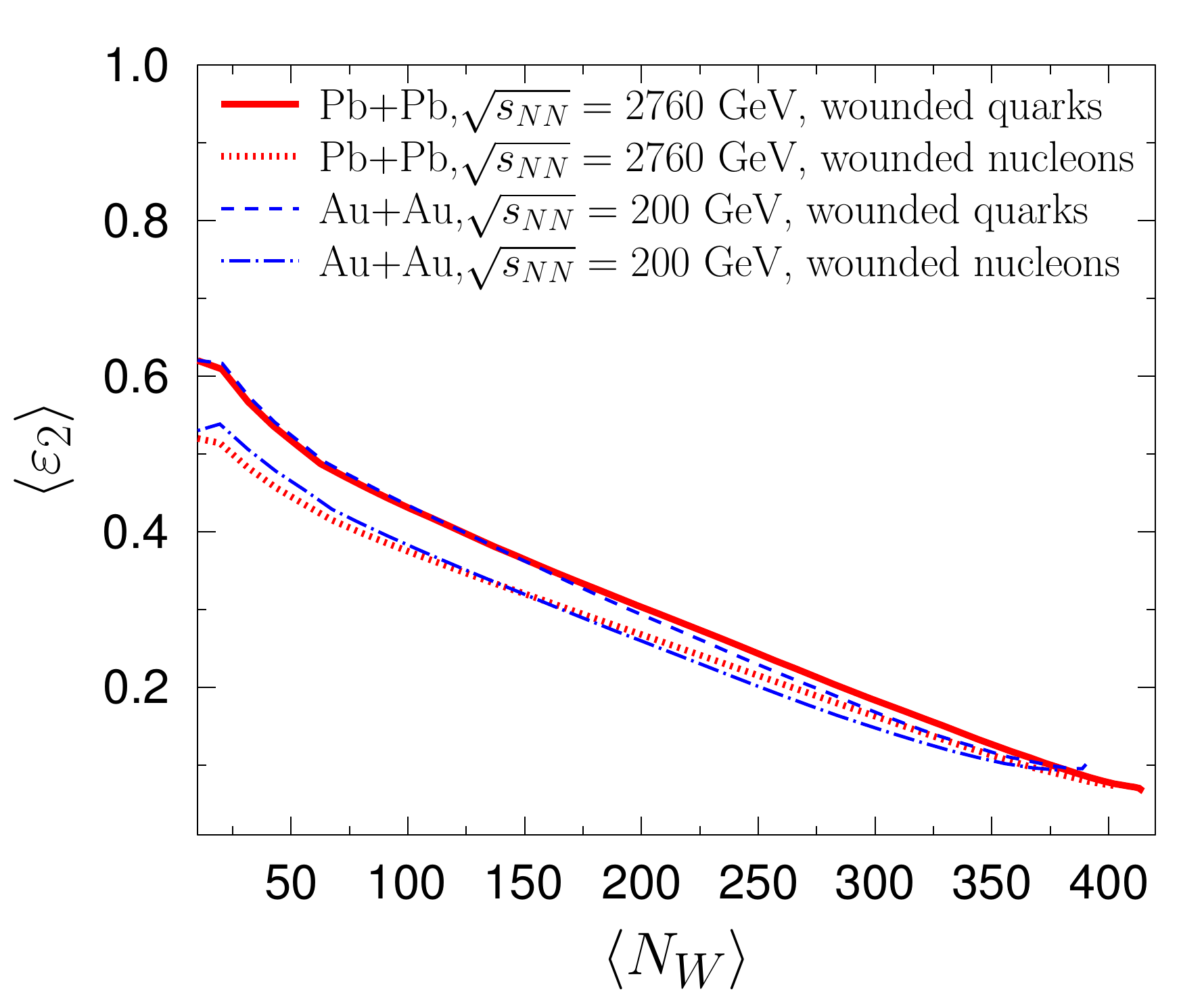} 
\end{center}
\vspace{-5mm}
\caption{Average ellipticity vs centrality for two selected reactions, evaluated in the wounded quark and wounded  nucleon models.  \label{fig:eps}}
\end{figure}

\begin{figure}
\begin{center}
\includegraphics[width=0.45 \textwidth]{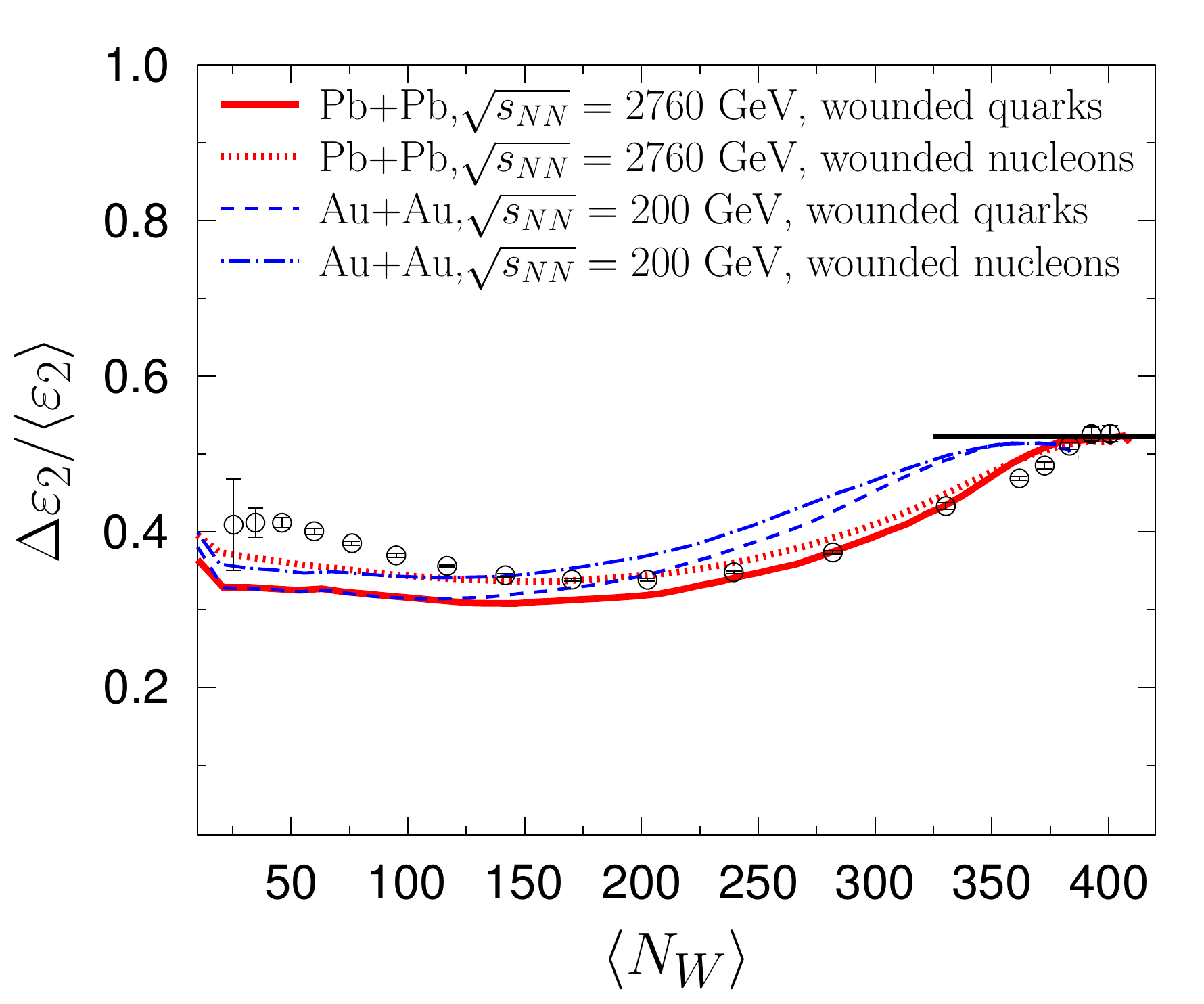} 
\end{center}
\vspace{-5mm}
\caption{The fluctuation measure $\Delta \varepsilon_{2}/\langle \varepsilon_{2} \rangle$ vs centrality, evaluated in the wounded quark and wounded nucleon models and compared to the data 
from ATLAS Collaboration~\cite{Aad:2013xma} (circles).   \label{fig:sigeps}}
\end{figure}

A basic feature of phenomenology of relativistic heavy-ion collisions is the development of harmonic flow due to geometry~\cite{Ollitrault:1992bk} and to event-by-event 
fluctuations~\cite{Alver:2006wh,Voloshin:2006gz,Hama:2007dq,Alver:2010gr,Luzum:2011mm,Bhalerao:2014xra}. In this section 
we compare predictions of the wounded nucleon and wounded quark models for the eccentricities of the initial state.
For the wounded nucleon case we use the mixing parameter $\alpha=0.145$ for $\sqrt{s_{NN}}=200$~GeV and 
$\alpha=0.15$ for $\sqrt{s_{NN}}=2.76$~TeV. The Gaussian smearing parameter is $\sigma=0.4$~fm for wounded nucleons and $\sigma=0.2$~fm for wounded quarks
(the value of $\sigma$ has tiny effects in the A+A collisions).
We note from Fig.~\ref{fig:eps} that the wounded quarks lead to larger ellipticity compared to the wounded nucleon case, with the effect 
reaching about 15\% at the peripheral collisions. 

We have checked that for the most central collisions the ratio 
$\varepsilon_3/\varepsilon_2$ approaches the limit of the independent source model 
(cf. Eq.~(7.8) of Ref.~\cite{Broniowski:2007ft}, see also the discussion in Ref.~\cite{Blaizot:2014wba}), namely
\begin{eqnarray}
\frac{\varepsilon_3}{\varepsilon_2}=\frac{\langle \rho^2 \rangle}{\langle \rho^3 \rangle}\sqrt{\frac{\langle \rho^6 \rangle}{\langle \rho^4 \rangle}}, \label{eq:epsrat}
\end{eqnarray}
where $\langle \rho^n \rangle$ denote the moments of the initial density for collisions at at zero impact parameter in the transverse radius $\rho=\sqrt{x^2+y^2}$. 
Numerically, the ratio of Eq.~(\ref{eq:epsrat}) is very similar for the wounded nucleon and wounded quark models and is $\sim 1.1$. As is known, 
the proximity of the most central values of $\varepsilon_2$ and $\varepsilon_3$ leads to problems in the description of $v_2$ and $v_3$ in viscous 
hydrodynamics, where triangular flow is quenched more strongly than the elliptic flow \cite{Luzum:2012wu,*Rose:2014fba}. As a result, the predicted values of $v_3$ are significantly smaller than $v_2$, in 
contrast to the experiment~\cite{CMS:2013bza,*ATLAS:2012at}.

When switching from nucleon to quark participants two effects influence the entropy distribution in the fireball, with opposite effect on the eccentricities.
Firstly, additional fluctuations at subnucleonic
 scales appear from individual entropy deposition from wounded quarks. 
Secondly,  the entropy is spread around the nucleon-nucleon collision point 
among the different positions of individual
 wounded quarks in the colliding 
nucleons. The second effect can be estimated from the rms size of the fireball
in p+p collisions (Sect. \ref{sec:npp}); it is of the same order ($0.4$~fm)
 as the width of the Gaussian centered at the positions of the wounded 
nucleons. The second effect comes in a similar way in both models and gives a
 reduction of the eccentricities. On the other, hand the first effect appears only when using subnucleonic degrees of freedom and leads to larger  eccentricities in the wounded quark model.
Correspondingly, as presented in Fig.~\ref{fig:sigeps}, scaled  event-by-event fluctuations of ellipticity are reduced with wounded 
quarks, which brings the results closer to the data in semi-central collisions. The horizontal line in  Fig.~\ref{fig:sigeps} 
corresponds to the limit of $\sqrt{4/\pi-1}$ reached in the  most central events~\cite{Broniowski:2007ft}. 

Results analogous to Figs.~\ref{fig:eps} and \ref{fig:sigeps} for the triangularity are very similar for the wounded quark and the wounded nucleon models, 
except for $\langle \varepsilon_3 \rangle$ at peripheral collisions, where 
wounded quarks give higher values than wounded nucleons, up to about 15\%.

\begin{figure}
\begin{center}
\includegraphics[width=0.44 \textwidth]{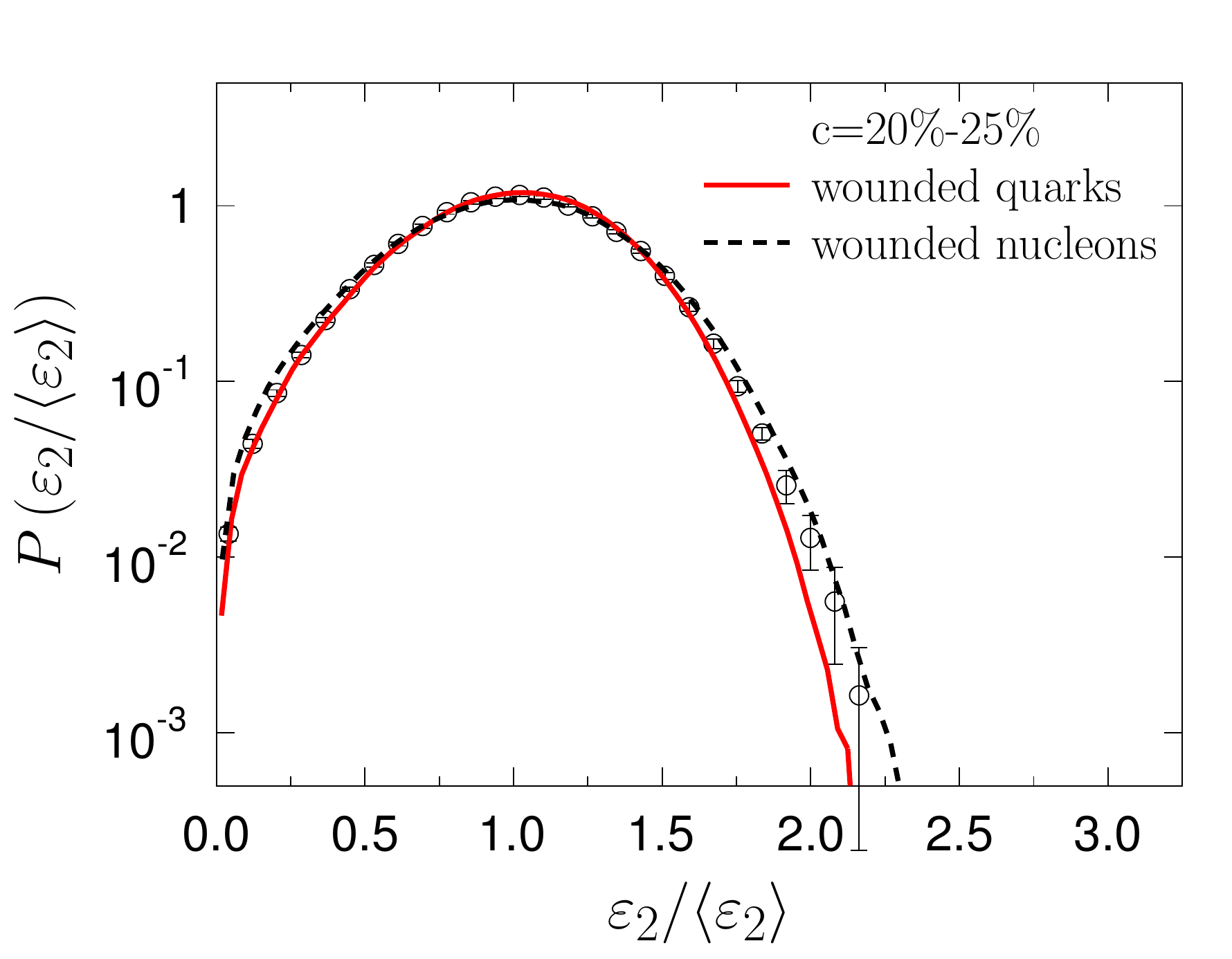} 
\end{center}
\vspace{-5mm}
\caption{Distribution of  $\varepsilon_{2}/\langle \varepsilon_{2} \rangle$  for centrality 20-25\%, compared to the data for 
$v_{2}/\langle v_{2} \rangle$ from ATLAS Collaboration~\cite{Aad:2013xma} (circles).  \label{fig:eps2}}
\end{figure}

\begin{figure}
\begin{center}
\includegraphics[width=0.44 \textwidth]{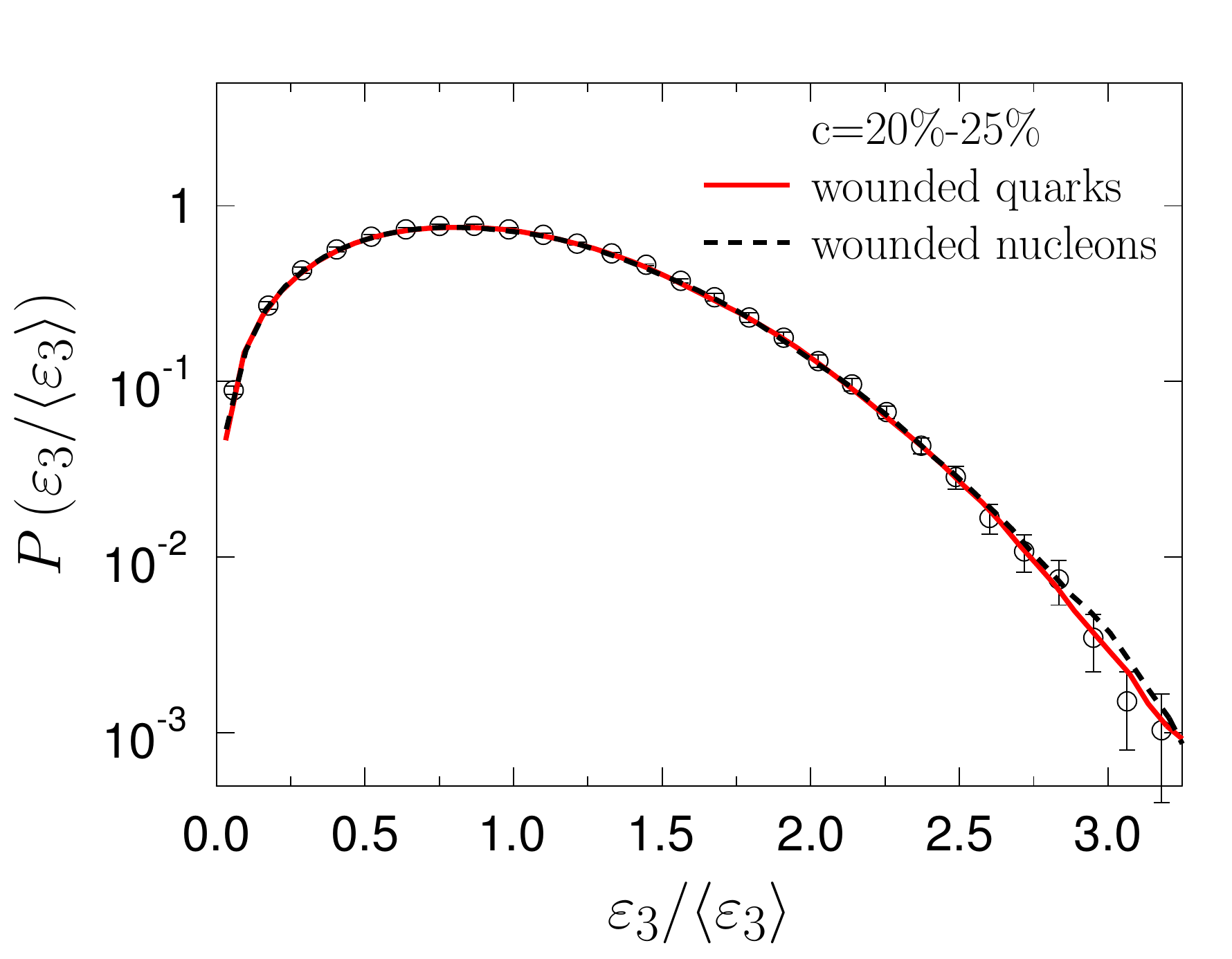}
\end{center}
\vspace{-5mm}
\caption{Same as in Fig.~\ref{fig:eps2} but for  the triangularity $\varepsilon_{3}$.  \label{fig:eps3}}
\end{figure}

In Figs.~\ref{fig:eps2} and \ref{fig:eps3} we compare the distributions of  $\varepsilon_{n}/\langle \varepsilon_{n} \rangle$ for ellipticity ($n=2$) 
and triangularity ($n=3$) for the two considered approaches. We note that the results are close to each other, with somewhat smaller tails for the case
of the wounded quarks (solid lines) for $n=2$, which moves 
the model a bit further from the data compared the the case of 
the wounded nucleons (dashed lines). 

\begin{figure}
\begin{center}
\includegraphics[width=0.45 \textwidth]{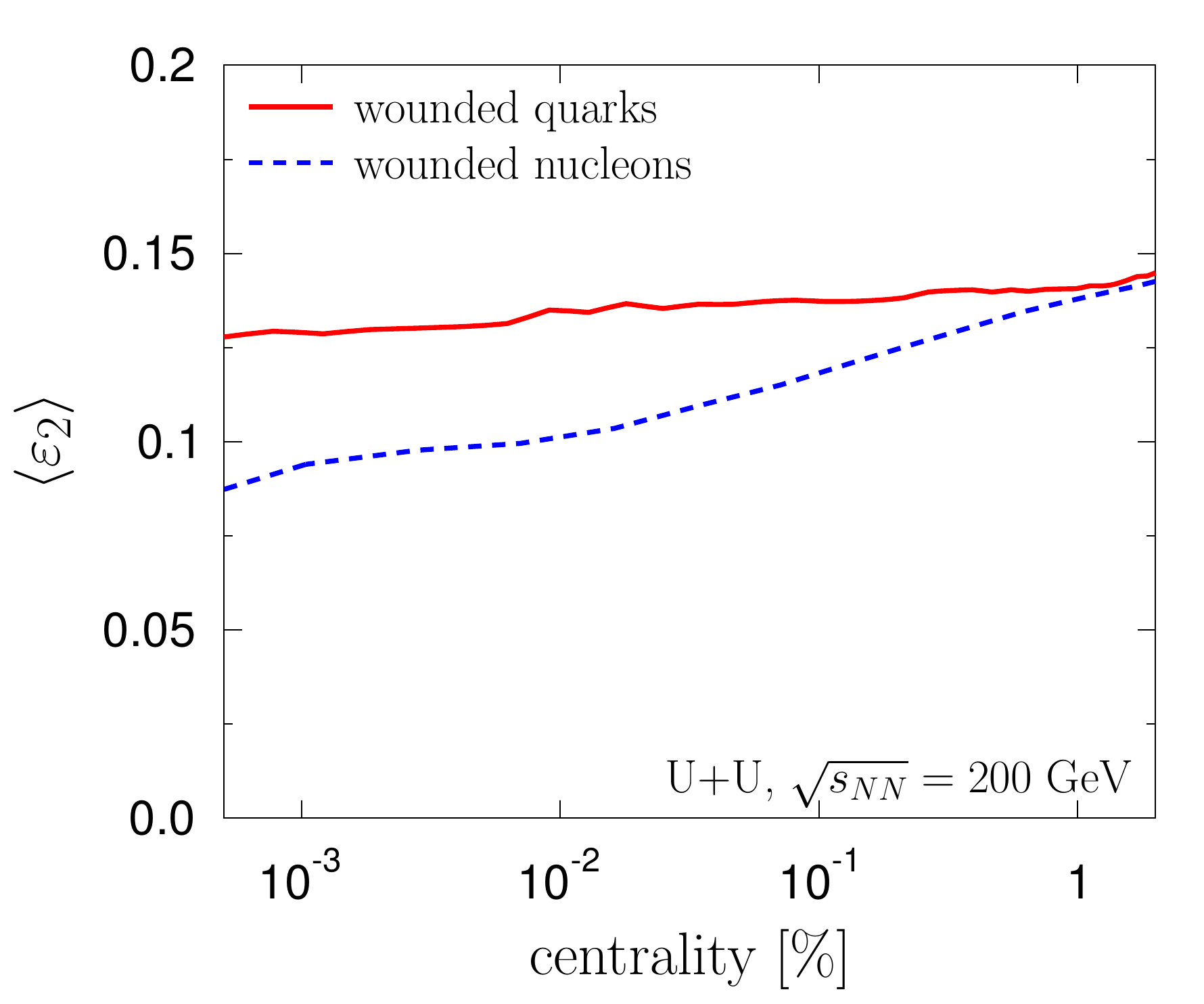} 
\end{center}
\vspace{-5mm}
\caption{Ellipticity $\varepsilon_2$ for the $^{238}$U+$^{238}$U collisions at $\sqrt{s_{NN}}=200$~GeV for the wounded quark (solid line) and wounded nucleon (dashed line) models,
plotted for centralities less than~1\%.  \label{fig:hyper}}
\end{figure}

A long-standing problem for the Glauber-based simulations were the results for the  $^{238}$U+ $^{238}$U collisions, measured at RHIC at $\sqrt{s_{NN}}=200$~GeV~\cite{Adamczyk:2015obl}. 
It had been expected that due to the intrinsic prolate deformation of the $^{238}$U nucleus, the results of the most central collisions  for the ellipticity should exhibit a knee structure \cite{Voloshin:2010ut}. It was later 
argued~\cite{Rybczynski:2012av} that this structure may be washed out
with large fluctuations of the overlaid distribution. Also, as shown recently, in a Glauber approach with {\em shadowing}~\cite{Chatterjee:2015aja}
the model predictions compare well to the data. Here we compare the predictions of the wounded quark and 
wounded nucleon models. It is evident from Fig.~\ref{fig:hyper} that at hyper-central collision the wounded nucleon model leads to fall-of of ellipticity with 
decreasing centrality $c$, whereas wounded quarks give a flattening, i.e., no knee behavior, in qualitative accordance to the data
and the wounded quark model studies reported in~\cite{Adamczyk:2015obl}.

\subsection{Size fluctuations}

In Ref.~\cite{Broniowski:2009fm} it was proposed that  transverse size fluctuations of the initial state lead to transverse momentum 
fluctuations. The mechanism is based on the simple fact that more compressed matter, as may happen from statistical fluctuations in the 
initial state, leads to more rapid radial hydrodynamic expansion, which then provides more momentum to hadrons. The mechanism was later tested with 
3+1~dimensional viscous event-by-event hydrodynamics~\cite{Bozek:2012fw}, with a somewhat surprising result that the effect leads to even larger (by about 30\%) fluctuations than
needed to explain the experimental data~\cite{Adler:2003xq,Adams:2005ka}, in particular for most central collisions. 
The basic formula for the mechanism of Ref.~\cite{Broniowski:2009fm} is
that the event-by-event scaled standard deviation of the average transverse momentum is proportional to the corresponding quantity for the transverse size, $\langle r \rangle$, of the
initial state:
\begin{eqnarray}
 \frac{\Delta\langle r \rangle}{\langle \langle r \rangle \rangle}= \beta  \frac{\Delta\langle p_T \rangle}{\langle \langle p_T \rangle \rangle},
\end{eqnarray}
where $\langle \langle . \rangle \rangle$ denotes the event-by-event average of the quantity averaged in each event, $\Delta$ denotes the standard deviation, 
and the measure of the transverse size is
\begin{eqnarray}
\langle r \rangle^2 = \frac{\int dx\,dy\, s(x,y) (x^2+y^2)}{\int dx\,dy\, s(x,y)}. \label{eq:rd}
\end{eqnarray}
The function $s(x,y)$ is the entropy density in the transverse coordinates $(x,y)$. The constant $\beta \sim 0.3$ depends on the hydrodynamic response, but not on the centrality of the collision, which allows 
for predictions. We introduce the variable 
\begin{eqnarray}
\langle r \rangle = \sqrt{\langle r \rangle^2},  \label{eq:r}
\end{eqnarray}
which is analyzed event by event, in particular $\langle \langle r \rangle \rangle$ is the event-by-event average of the size of Eq.~(\ref{eq:r}). 

\begin{figure}
\begin{center}
\includegraphics[width=0.42 \textwidth]{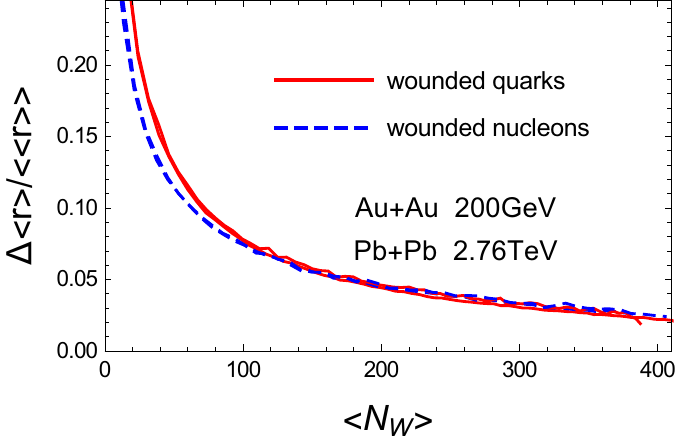} 
\end{center}
\vspace{-5mm}
\caption{Scaled size fluctuations of the initial fireball, plotted as a function of centrality evaluated in the wounded quark and nucleon models with an overlaid gamma 
distribution. The results for Au+Au at $\sqrt{s_{NN}}=200$~GeV and Pb+Pb at $\sqrt{s_{NN}}=2.76$~TeV overlap. \label{fig:size}}
\end{figure}

The results for the event-by-event scaled standard deviation of the size are presented in Fig.~\ref{fig:size}. 
We compare the wounded nucleon (\ref{eq:wn}) and the wounded quark (\ref{eq:wq}) models with an overlaid gamma distribution.
We note that at low centralities the size fluctuations in both models are very close for the central collisions, whereas for peripheral collisions the size 
fluctuations are larger in the wounded quark model. The results for the two sample reactions, Au+Au at $\sqrt{s_{NN}}=200$~GeV and Pb+Pb at $\sqrt{s_{NN}}=2.76$~TeV, are virtually 
indistinguishable, as the corresponding curves lie on top of each other.

\section{${\mathbf p}$+A collisions \label{sec:pA}}

Ultrarelativistic p+A collisions are an important testing ground for theoretical approaches, due to the
expected onset of collectivity~\cite{Bozek:2011if,Bozek:2012gr,Bozek:2013uha,Bozek:2013ska}.
It is thus important to address this issue with the model based on wounded quarks, as we search a uniform description 
of particle production. In this section we present the results for hadron multiplicity, the fireball size, and eccentricities.

\subsection{Multiplicity distribution \label{sec:mult_pA}}

\begin{figure}
\begin{center}
\includegraphics[width=0.42 \textwidth]{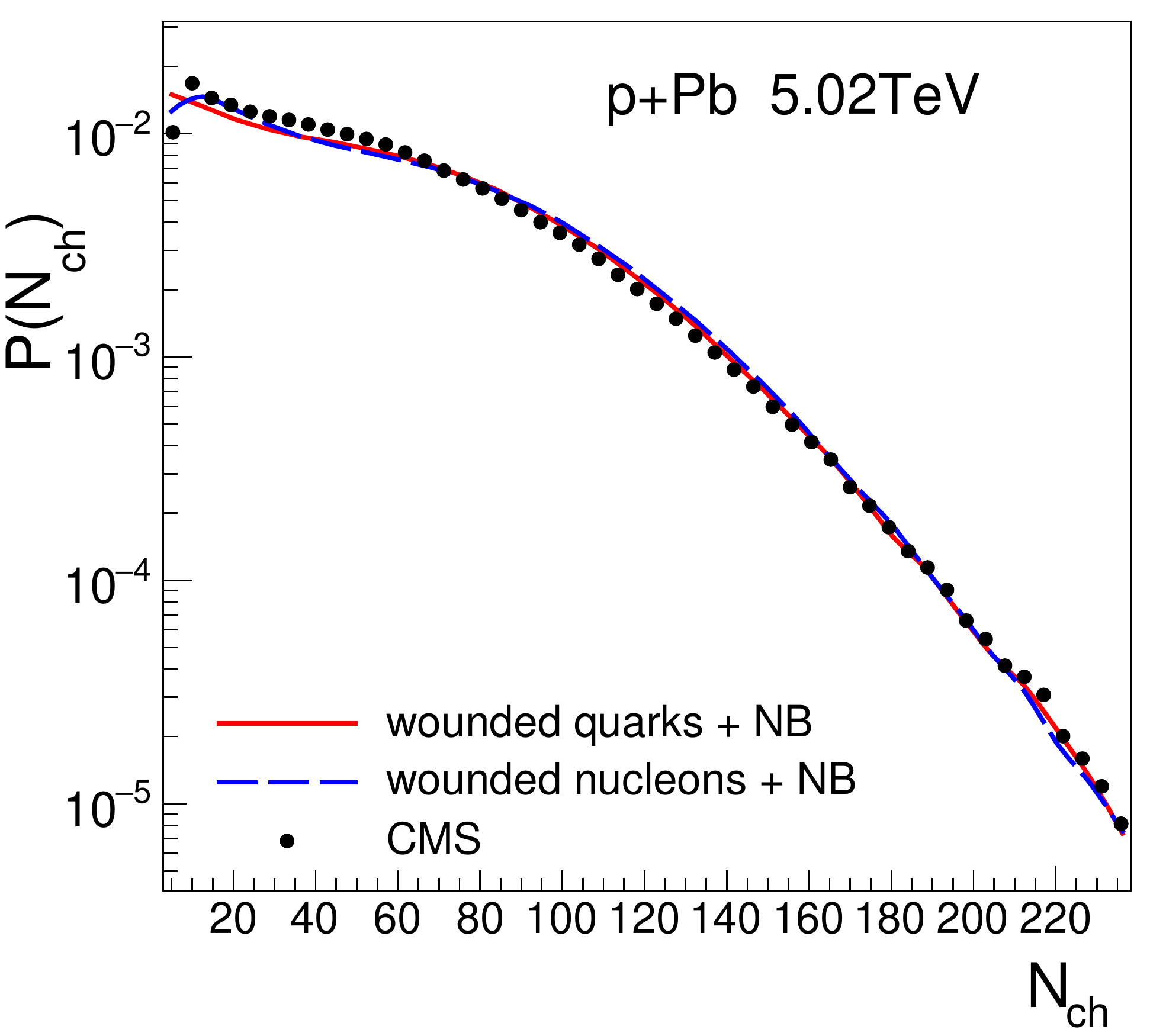} 
\end{center}
\vspace{-5mm}
\caption{Multiplicity of charged hadrons in p+Pb collisions from the wounded quark and wounded nucleon models, compared to the 
preliminary CMS data~\cite{cmswiki}. Appropriate negative binomial distribution is overlaid over the distribution of sources (see text). \label{fig:npPb}}
\end{figure}

To reproduce the experimental hadron multiplicity distribution in p+A collisions, in particular the tail 
at very high multiplicities, one needs to overlay an additional distribution over the 
Glauber sources, as explained in Sec.~\ref{sec:over}. The physical meaning of this procedure is that  sources deposit 
entropy at a varying strength. The mechanism was already proposed in the original wounded nucleon model~\cite{Bialas:1977en}. 
Experimental data are well reproduced with an overlaid negative binomial distribution (Sec.~\ref{sec:over}).
In Fig.~\ref{fig:npPb} we compare the wounded-nucleon and the wounded-quark model predictions, compared to the CMS data~\cite{cmswiki}. The
high-multiplicity tail is properly reproduced when the $\kappa$ parameter of the negative binomial distribution is about 0.5. 
In Fig.~\ref{fig:npPb} we use $\kappa=0.54$ for the wounded quark model, whereas $\kappa=0.9$ for the wounded nucleon model ($\alpha=0$ in Eq.~(\ref{eq:wn})). 
Here we follow the convention that the parameters of the negative binomial distribution correspond to overlaying over the nucleons and not the nucleon pairs, as implied by Eq.~(\ref{eq:wn}). 
The parameter $\bar{n}$ is adjusted such that mean experimental multiplicities
are reproduced at a given experimental acceptance and efficiency, namely $\bar{n}=3.9$ for the wounded quark and $\bar{n}=6.2$ for wounded nucleon.

We note a fair description of the tail of the distributions in Fig.~\ref{fig:npPb}, whereas, admittedly, there are departures at low values of $N_{\rm ch}$.  

In Fig.~\ref{fig:npPb} the coordinate $N_{\rm ch}$ corresponds to the number of tracks in the CMS detector. Unfortunately, no hadron multiplicity data corrected for acceptance and efficiency are available, 
such that at present we cannot overlay the p+Pb point on other results in Fig.~\ref{fig:dndeta_scaled}.

\subsection{Fireball in p+Pb}

\begin{figure}
\begin{center}
\includegraphics[width=0.515 \textwidth]{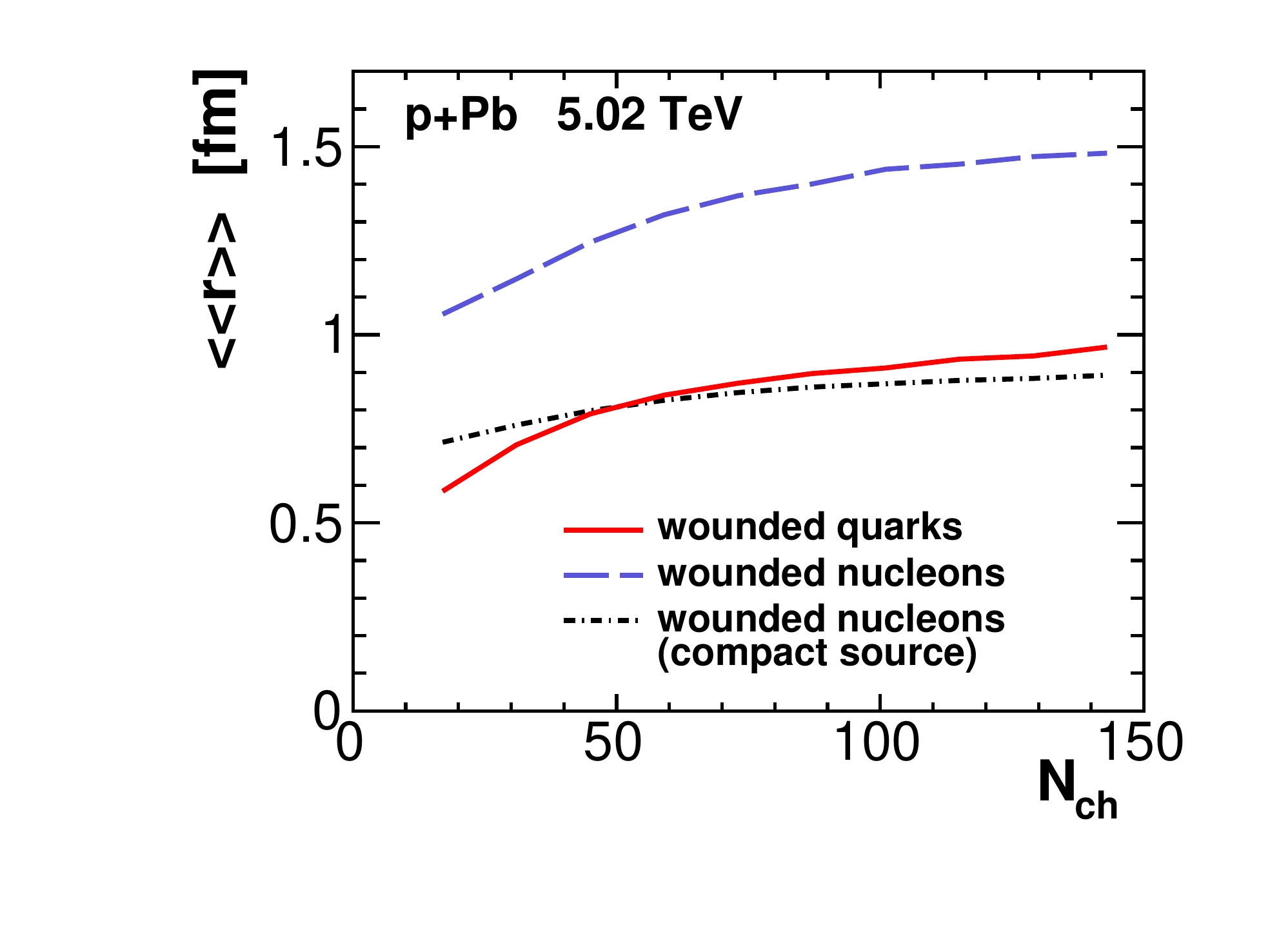} 
\end{center}
\vspace{-14mm}
\caption{Initial size of the fireball in p+Pb collisions at $\sqrt{s_{NN}}=5.02$~TeV evaluated from Eq.~(\ref{eq:r}) and 
plotted against the multiplicity of produced charged hadrons. The solid line corresponds to the wounded quark model with the smearing parameter $\sigma=0.2$~fm, and is compared to
wounded nucleons with $\sigma=0.4$~fm in the standard form (dashed line) and the {\em compact source} version (dot-dashed line).  \label{fig:size_pPb}}
\end{figure}

The size and the shape of the fireball are important characteristics of the initial state. The size of the fireball determines the transverse push and
interferometry correlations~\cite{Bozek:2013df,*Schenke:2014zha}, whereas
the eccentricities generate  harmonic
flow coefficients in the collective expansion. In the hydrodynamic model the best description of the data is obtained when
the entropy is deposited in between the two colliding nucleons, in the so called {\em compact source} scenario~\cite{Bzdak:2013zma,Bozek:2013uha}, where the entropy
is deposited in the middle between the two colliding nucleons. The prescription corresponds to Eq.~(\ref{eq:wn}) with $\alpha=1$.
The effective  number of sources is $N_{\rm W}-1$ in this case, and the parameters of the overlaid binomial distribution fitted to the experiment 
as in Fig.~\ref{fig:npPb} are $\bar{n}=6.6$ and $\kappa=1.0$.

In Fig.~\ref{fig:size_pPb} we show the size measure of Eq.~({\ref{eq:r}) plotted as a function of the number of produced charged hadrons. We note that the results from the wounded quark model are closer 
to the {\em compact source} variant of the wounded 
nucleon model than to its standard version. 
Therefore the wounded quark model gives a microscopic motivation for the prescription used in the {\em compact source} scenario~\cite{Bozek:2013uha}.

In Fig.~\ref{fig:eps_pPb} we present the ellipticity and the triangularity
in the wounded quark model. The deformation of the initial fireball is similar 
as in the standard wounded nucleon model~\cite{Bozek:2013uha}. We expect that the 
shape and  size of the fireball obtained in the wounded quark model 
is a good initial condition for the hydrodynamic description of  measurements made in  p+Pb collisions.

\begin{figure}
\begin{center}
\includegraphics[width=0.515 \textwidth]{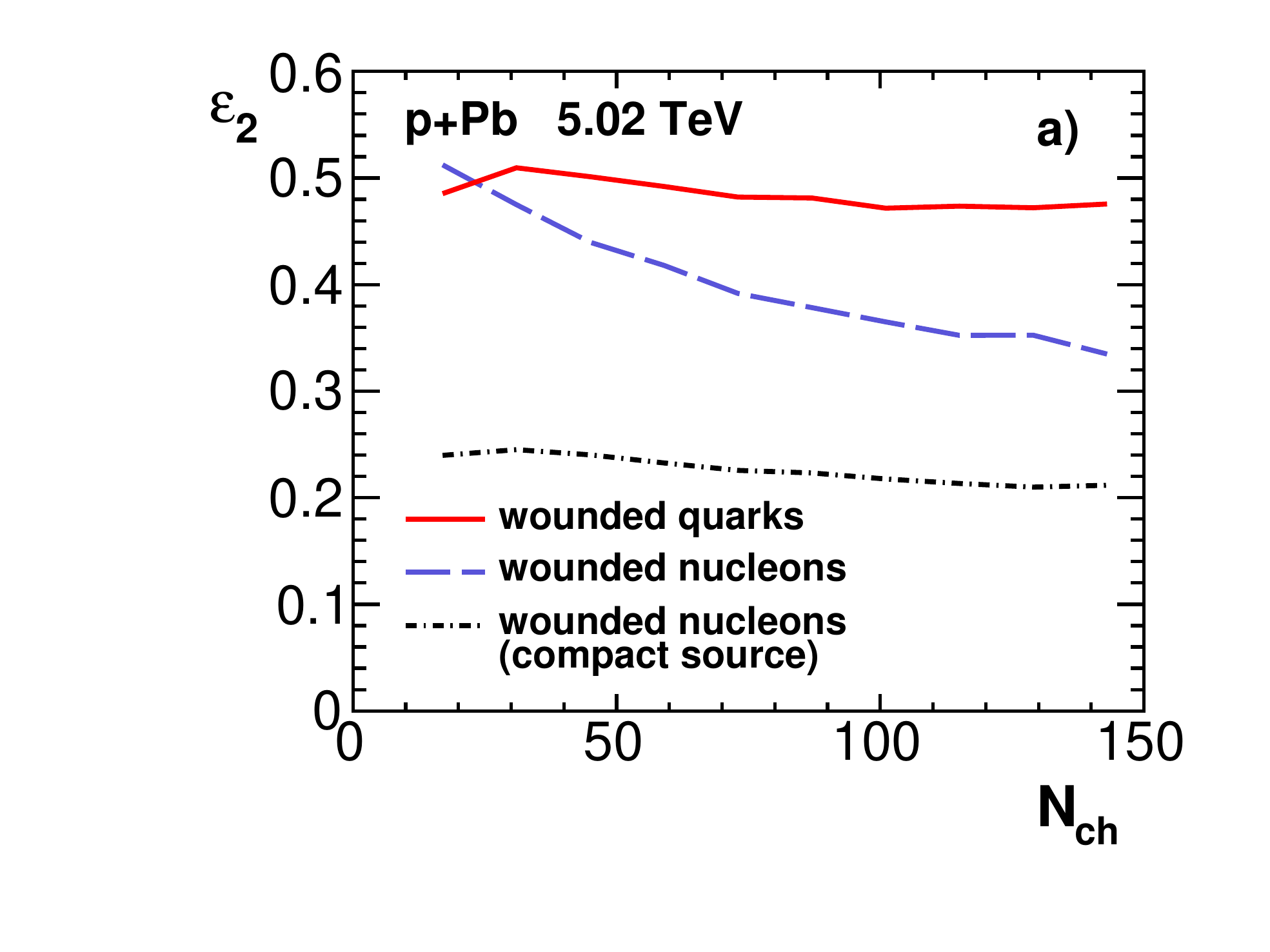} 
\end{center}
\vspace{-24mm}
\begin{center}
\includegraphics[width=0.515 \textwidth]{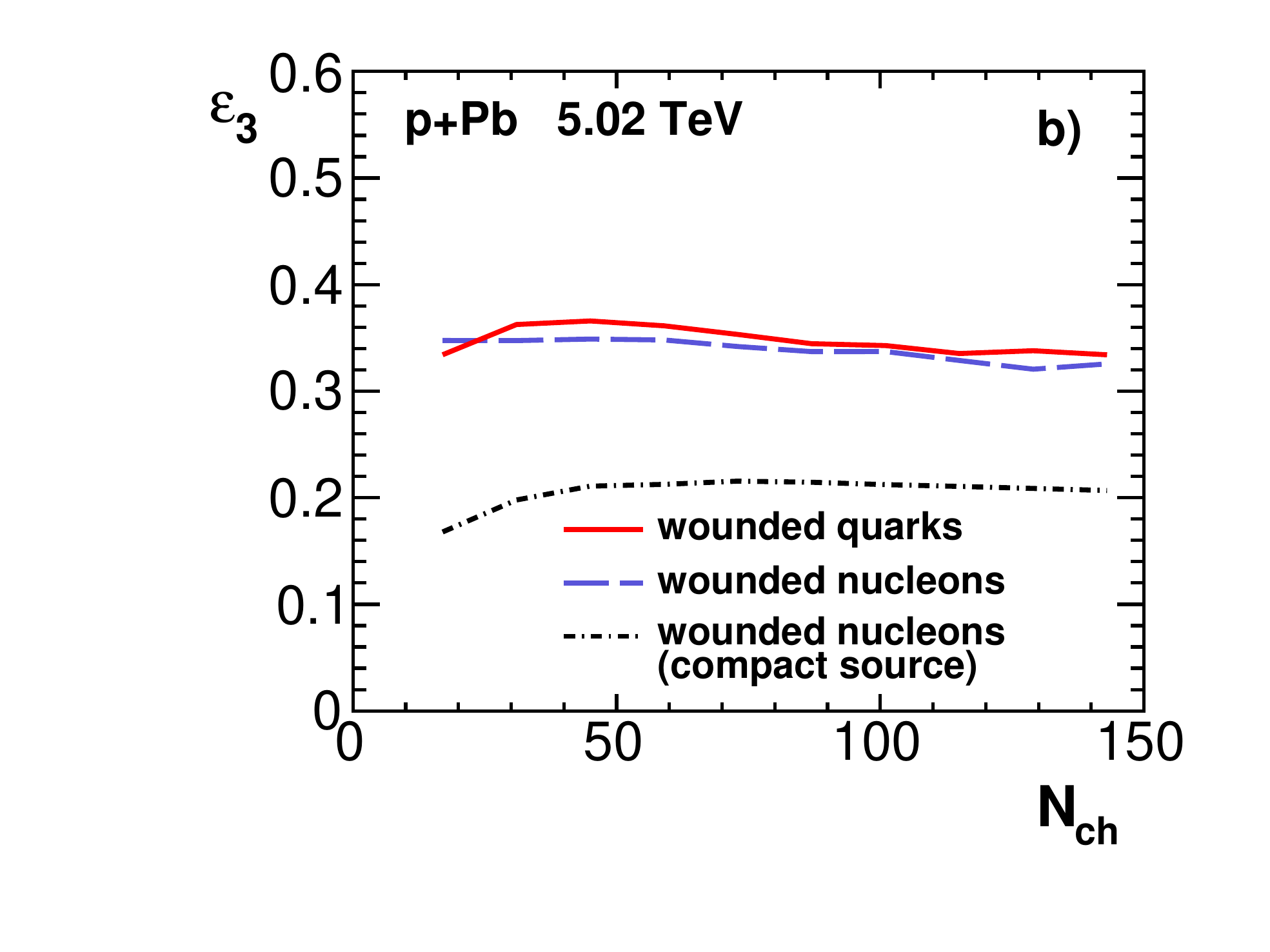} 
\end{center}
\vspace{-12mm}
\caption{Ellipticity (panel a) and triangularity (panel b) of the initial fireball in p+Pb collisions at $\sqrt{s_{NN}}=5.02$~TeV from the wounded quark model
with smearing parameter $\sigma=0.2$~fm, plotted as functions of the multiplicity of produced charged hadrons.  \label{fig:eps_pPb}}
\end{figure}

\section{${\mathbf p}$+${\mathbf p}$ collisions  \label{sec:npp}}

With subnucleonic degrees of freedom we have an opportunity to analyze the p+p collisions.
The proton-proton inelastic collision profile  in the wounded quark model is described using  quark-quark collisions 
(Sec.~\ref{sec:qconst} and Appendix~\ref{sec:inel}). The total inelastic cross section at each energy is reproduced
with an energy dependent  quark-quark cross section.

\subsection{Multiplicity distributions}

The mean number of wounded quarks in p+p collisions 
increases mildly with the energy, as we find $Q_{\rm W}=2.6$, $2.75$, and $2.81$ at $\sqrt{s}=200$~GeV, 
$2.76$~TeV, and $7$~TeV  respectively. The multiplicity of charged hadrons comes from a convolution of the
multiplicity distributions of particles produced from each wounded quark and the distribution of the number
of wounded quarks (Sec. \ref{sec:over}). 

\begin{figure}
\begin{center}
\includegraphics[width=0.47 \textwidth]{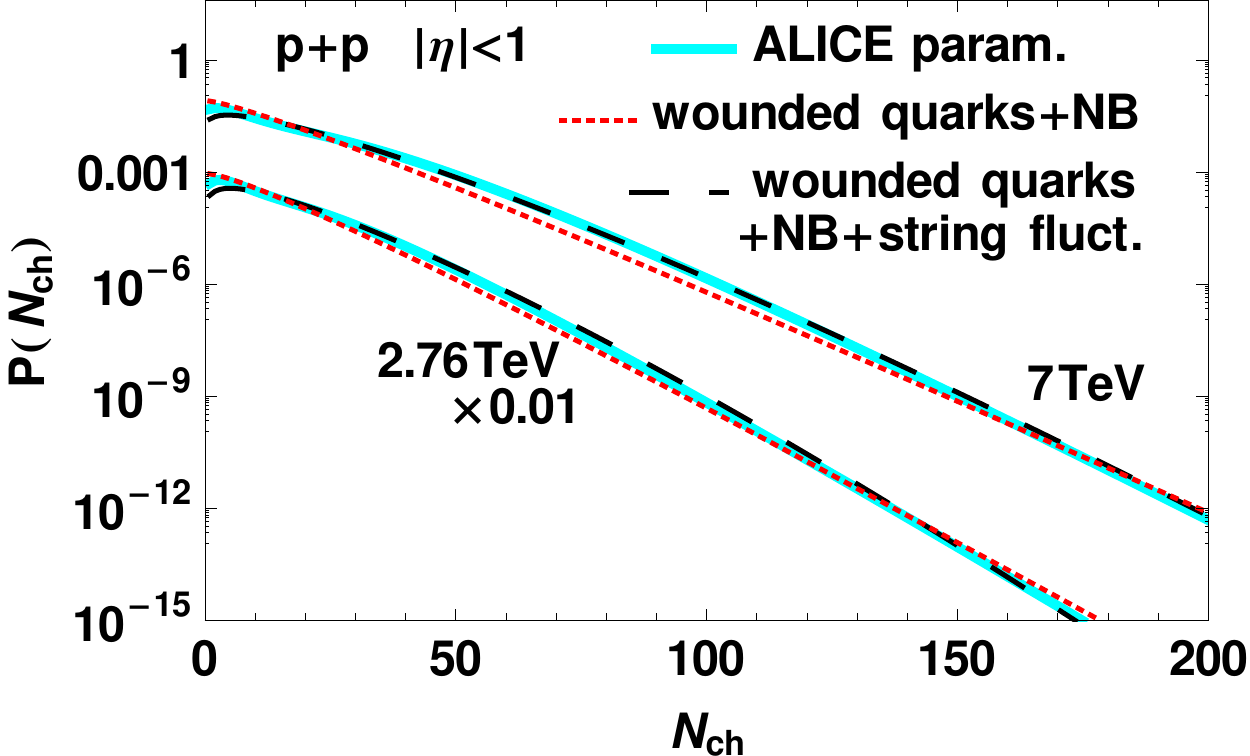} 
\end{center}
\vspace{-5mm}
\caption{Charged particle multiplicity distribution for $|\eta|<1$. Solid lines denote a
parametrization of experimental data from the ALICE Collaboration~\cite{Adam:2015gka} 
for $\sqrt{s}=7$~TeV (upper curves) and for $\sqrt{s}=2.76$~TeV (lower curves, multiplied by 0.01).
The multiplicity distribution from the wounded quark model convoluted with a negative binomial distribution is 
denoted with dotted lines, whereas dashed lines represent the calculation where both string fluctuations in rapidity 
and a negative binomial distribution are convoluted for each wounded quark. \label{fig:npp}}
\end{figure}

In Fig.~\ref{fig:npp} we show the multiplicity distribution of charged hadrons (for the acceptance window $|\eta|<1$)
in inelastic p+p  collisions at two energies: 
$2.76$ and $7$~TeV. The experimental results are represented by a numerical parametrization using 
a sum of two negative binomial distributions~\cite{Adam:2015gka}.
These curves for the multiplicity distributions are reproduced in the wounded quark model by adjusting
the parameters of the negative binomial distribution convoluted with the wounded quark distribution.
We find  $\bar{n}=3.3$ and $\kappa=0.52$ at $7$~TeV, and  
 $\bar{n}=2.8$ and $\kappa=0.55$ at $2.76$~TeV (dotted lines in Fig.~\ref{fig:npp}).
The agreement of our model with the ALICE phenomenological fit, while not perfect, holds approximately over 12 orders of magnitude.  
 
The scenario discussed above assumes that all the wounded quarks contribute to  particle production in the considered pseudorapidity interval. 
Alternative scenarios are possible, where each wounded quark contributes to particle production  in a limited pseudorapidity interval. An example 
of such a case is the flux tube model, where strings decay into particles in a rapidity interval limited by the rapidity of the leading 
charges~\cite{Andersson:1983ia,Bozek:2015bna,Monnai:2015sca}. Without entering into details of a particular model of the random deposition of
 entropy in rapidity, we make a simple estimate using an extreme scenario. Namely, a wounded quark contributes
 (or not)  with probability $1/2$ to hadron production in the central rapidity interval. It is a scenario assuming largest possible fluctuations from a flux-tube mechanism. The goal of this exercise is to show that in both extreme scenarios,
 equal, smooth in rapidity contribution from each wounded quark and strongly fluctuating contribution from each quark, the observed hadron multiplicity distribution can be reproduced. 
In the model  with fluctuating contribution from each wounded quark, the parameters of the negative binomial distribution are
$\bar{n}=7.9$ and $\kappa=1.7$ at $7$~TeV, and  $\bar{n}=7.4$ and $\kappa=2.2$ at $2.76$~TeV (dashed lines in Fig. \ref{fig:npp}). These estimates 
show that in the wounded quark model there is a possibility
to accommodate for a mechanism involving additional fluctuations in the energy deposition in a fixed rapidity
interval, with a proper adjustment of the parameters of the convoluted negative binomial distribution.

\subsection{Fireball in p+p}

\begin{figure}
\begin{center}
\includegraphics[width=0.515 \textwidth]{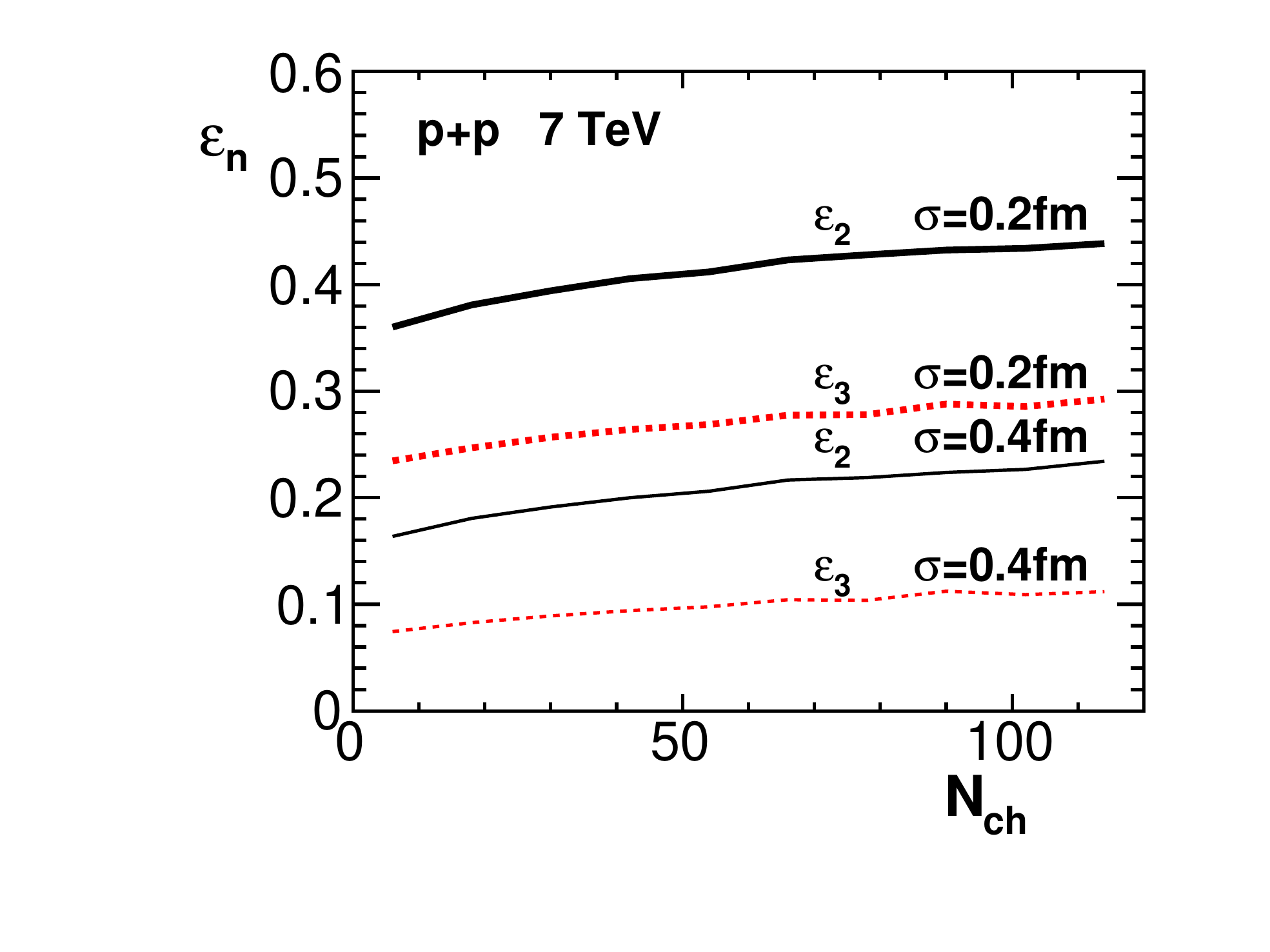} 
\end{center}
\vspace{-15mm}
\caption{Ellipticity $\varepsilon_2$ (solid lines) and triangularity $\varepsilon_3$ (dotted lines) 
of the fireball in p+p collisions at $\sqrt{s}=7$~TeV with the smoothing width $\sigma=0.2$~fm (thick lines) and 0.4 fm
(thin lines), as a function of the  mean charged multiplicity for $|\eta|<2.4$. \label{fig:e23pp}}
\end{figure}

In an inelastic p+p collisions two or more wounded quarks take part in the collision. The distribution of the wounded quarks in the transverse plane generates transverse shape deformations of the initial fireball.
In such a  small system it is essential to check the sensitivity to 
the range of the deposition of entropy from each quark.
We take a Gaussian profile (\ref{eq:smooth}) of width of $0.2$ 
or $0.4$~fm. The distribution in the transverse plane takes into account the fluctuations in the
 entropy deposition from each wounded quark, with an overlaid gamma distribution with parameters
 adjusted to the multiplicity distribution as explained above. The total entropy in the fireball is rescaled to the mean
 multiplicity for $|\eta|<2.4$. The eccentricities show a very weak dependence on multiplicity (Fig.~\ref{fig:e23pp}).
We note that the triangularity is much smaller than the ellipticity. In a scenario with a collective 
expansion in p+p collisions this would  give $v_3$ much smaller than $v_2$. The smoothing parameter of the Gaussian in 
the initial fireball has a decisive role in determining the magnitude of the eccentricity.

\begin{figure}
\begin{center}
\includegraphics[width=0.42 \textwidth]{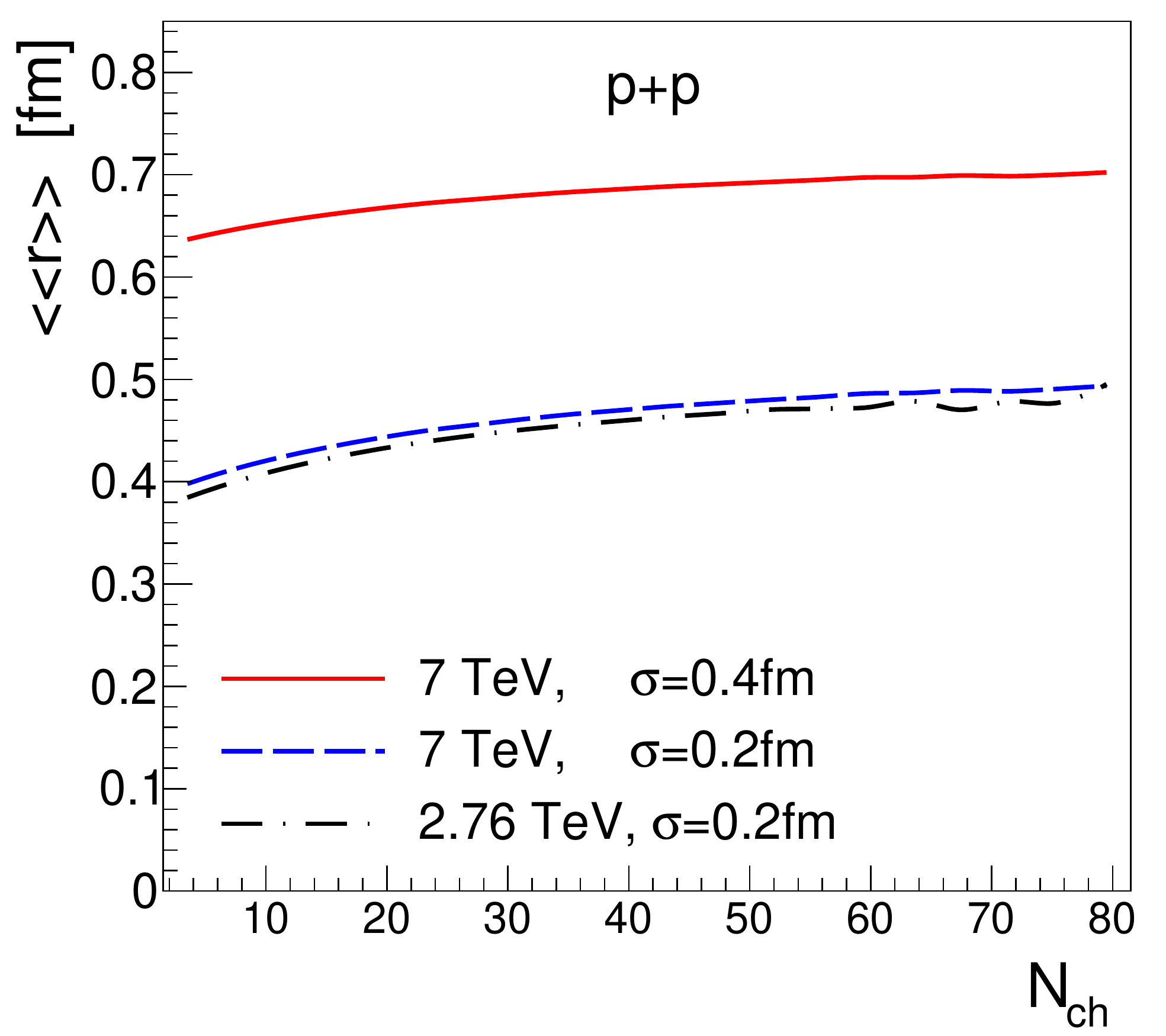} 
\end{center}
\vspace{-5mm}
\caption{Initial size of the fireball in p+p collisions at the LHC energies evaluated with different Gaussian smearing parameters and 
plotted against the multiplicity of produced charged hadrons. \label{fig:size_pp}}
\end{figure}

In Fig.~\ref{fig:size_pp} we display the average size of the system formed in p+p collisions at two LHC energies and for two values of the 
Gaussian smearing parameters. This observable reflects the size of the proton, the quark-quark inelasticity profile, and the Gaussian smearing parameter.
We note that the size, especially for the case of $\sigma=0.4$~fm, is not much smaller than in the p+Pb system 
shown in Fig.~\ref{fig:size_pPb}. This opens the opportunity of using the fireball from the wounded quark model in 
hydrodynamic calculations for p+p collisions~\cite{Bozek:2009dt,Werner:2010ny}.

\section{More partons \label{sec:more}}

The number of effective subnucleonic degrees of freedom in the nucleon
could be different from three, where quark constituents are assumed.
For instance, at the ISR energies the proton-proton scattering amplitude can be very well  described using
a quark-diquark model of the nucleon~\cite{Bialas:2006qf,CsorgO:2013kua}.
In the preceding sections we have assumed that the proton is composed 
of three quarks, as used in many other recent studies~\cite{Adler:2013aqf,Lacey:2016hqy,Mitchell:2016jio,Zheng:2016nxx}.
If protons are composed of numerous $N_p$ partons, proton-proton collisions
can be described in the Glauber optical model~\cite{d'Enterria:2010hd}.
In the optical limit, the parameter defining the inelastic collisions
is $N_p^2 \sigma_{pp}$, where $\sigma_{pp}$ is the parton-parton cross section.
Similarly, in the Monte Carlo Glauber model, 
when the number of partons in the proton 
increases, the parton-parton cross section decreases. A model with 
different numbers of partons in the proton was recently studied
by Loizides~\cite{Loizides:2016djv}, with a black-disc prescription for 
the parton-parton scattering. The cross section $\sigma _{pp}$ can then be 
adjusted to reproduce $\sigma_{NN}$ for each $N_p$. It was found that with
 a large   number of partons in a nucleon the increase of the number 
of wounded partons when going from peripheral to  central events is stronger. 

In our study we adjust the distribution profile of partons in the nucleon,
as well as $\sigma_{pp}$ and the parameters of the parton-parton inelastic scattering
profile to reproduce the inelasticity profile in p+p scattering, 
in the same way as  for $N_p=3$, as presented in Appendix~\ref{sec:inel}. 
We  find $\sigma_{pp}=22.96, \ 11.93,\ 7.32$, and $4.96$~mb for $N_p=2,\ 3,\ 4$, and $5$, respectively. 
The relation $ \sigma_{pp} \propto 1/N_p^2$ holds only approximately for the considered small number of partons in 
the nucleon.

\begin{figure}
\begin{center}
\includegraphics[width=0.42 \textwidth]{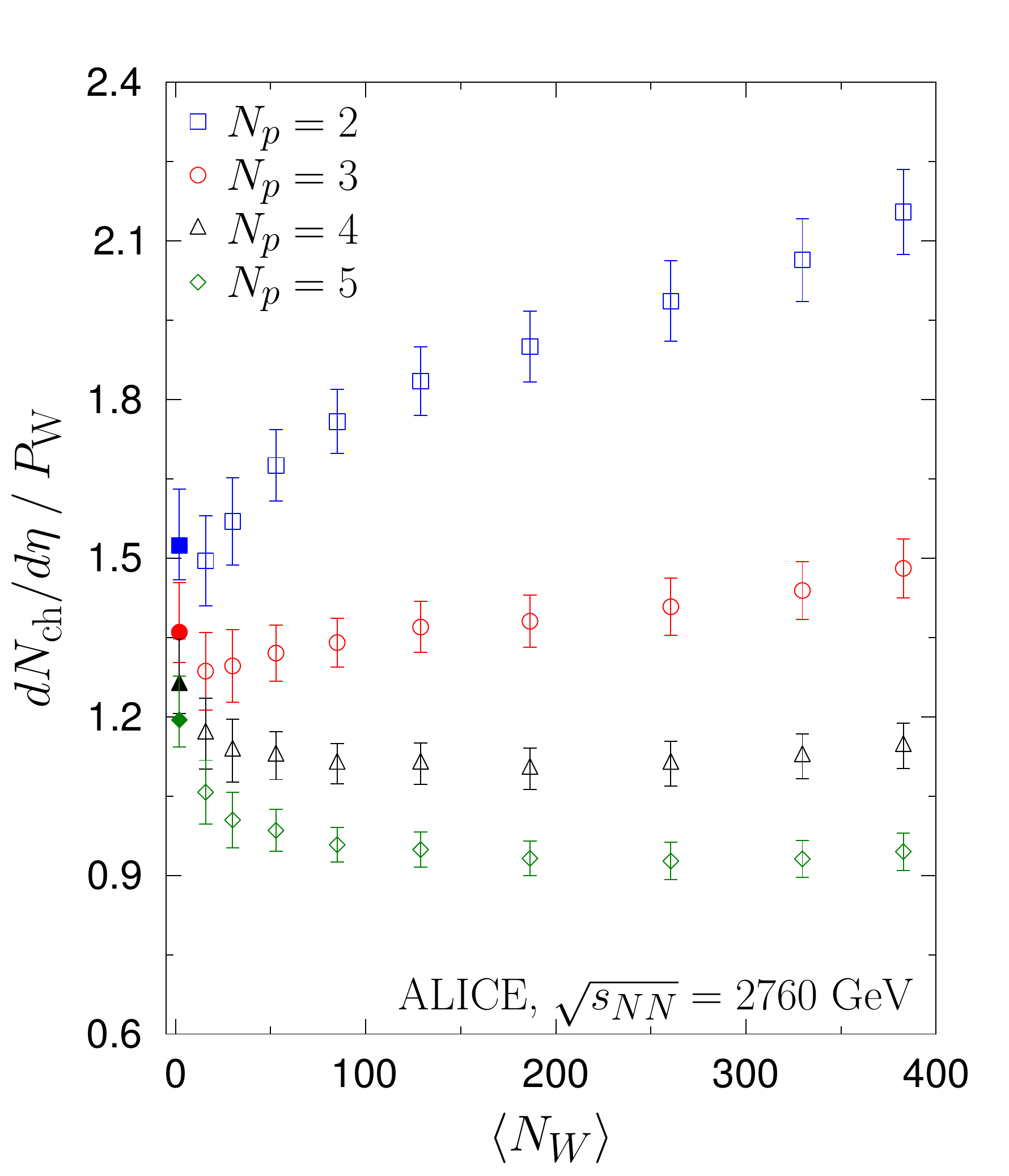} 
\end{center}
\vspace{-5mm}
\caption{The average multiplicity per unit of pseudorapidity,  $dN_{\rm ch}/d\eta$, divided by the number of wounded partons $P_{\rm W}$ for the Pb+Pb collisions at 
$\sqrt{s_{NN}}=2.76$~TeV (empty symbols), evaluated with different numbers of constituent quarks in the nucleon, $N_{\rm W}$.  The filled symbols at $\langle N_{\rm W} \rangle=2$
correspond to p+p collisions.
\label{fig:dndeta_nq}}
\end{figure}

The number of produced hadrons in A+A collisions is taken as proportional to the number
of wounded partons. The reduction of the cross section with 
the number of partons in the nucleon means that the number of 
wounded partons approaches the binary scaling in the Glauber model.
It means that with decreasing parton-parton cross section the number 
of produced hadrons as a function of centrality interpolates between
the wounded nucleon scaling and the binary scaling. This dependence 
can be used to estimate the effective number of partons involved 
in inelastic nucleon-nucleon collisions. The charged particle density in Pb+Pb
collisions divided by the number of partons is shown in Fig.~\ref{fig:dndeta_nq}.
We notice that the number that best describes the scaling of the particle
multiplicity is $3$ or slightly above, where the curve is flat as a function of centrality. 
This argument gives support to results for collisions at the LHC energies
 presented in the preceding 
sections, where $N_p=3$ was used. 

\section{Conclusions \label{sec:concl}}

We have explored in detail the Glauber Monte Carlo implementation of the wounded quark model, applied to 
particle production and for calculating 
the characteristics of the initial state in ultrarelativistic heavy-ion collisions. 
We have applied the model to p+p, p+A, d+A, He+A, Cu+Au, Au+Au, U+U, and Pb+Pb collisions at RHIC and the LHC energies, 
confirming an approximate linear scaling of the charged hadron density at midrapidity with the number 
of wounded quarks. 

To constrain the parameters of the effective subnucleonic degrees of freedom in the nucleon, we reproduce 
the proton-proton inelasticity profile in the impact parameter by adjusting 
the distribution of quarks in the nucleon and the quark-quark inelasticity profile. We find that 
the effective size of the proton increases weakly with the collision energy, whereas the  
growth of the quark-quark cross section is substantial and is responsible for the increase of the 
inelastic nucleon-nucleon cross section.
The multiplicity distributions in p+p and p+Pb collisions are used
to constraint the overlaid entropy fluctuations, here taken as an 
additional gamma distribution superposed over the distribution of the initial Glauber sources. 

Our conclusions are as follows:

\begin{enumerate}

 \item The production of particles at mid-rapidity follows the wounded quark scaling, with the quantity 
 $dN_{\rm ch}/d\eta / Q_{\rm W}$ displaying approximately a flat behavior with centrality (cf. Fig.~\ref{fig:dndeta_scaled}). This confirms the 
 earlier reports~\cite{Adler:2013aqf,Adare:2015bua,Lacey:2016hqy,Zheng:2016nxx} of the wounded quark scaling, but with phenomenologically motivated parameters 
of the quark-quark interaction.
At the LHC collision energy of  $\sqrt{s_{NN}}=2.76$~TeV the production per wounded quark in Pb+Pb collisions is compatible with 
the analogous quantity in p+p collisions, which is an essential feature for the consistency of the approach, displaying the universality 
of the particle production based on superposition of  individual collisions. 

\item At lower collision energies, such as $\sqrt{s_{NN}}=200$~GeV, the universality is far from perfect and the obtained scaling 
 is approximate, exhibiting some dependence on the reaction. Moreover, we note in Fig.~\ref{fig:dndeta_scaled} that the corresponding 
 p+p point is higher by about 30\% from the band of other reactions. This indicates that at lower energies there are corrections to the independent production 
 from the wounded quarks. Also, this may suggest a smaller number
of effective  subnucleonic degrees of freedom than three at these energies.

\item Average eccentricities and their event-by-event fluctuations show a similar dependence on centrality in the wounded  
quark model and in the wounded nucleon model amended with binary collisions of Eq.~(\ref{eq:wn}). The average eccentricities are somewhat larger in the wounded quark model 
compared to the wounded nucleon case, reflecting the presence of additional fluctuations at subnucleonic scales.

\item For  collisions of the deformed U+U nuclei at RHIC ($\sqrt{s_{NN}}=200$~GeV), we find that the average ellipticity in the wounded quark model  flattens at very central collisions, as opposed 
to the wounded nucleon case. This is qualitatively in agreement with the absence of the knee structure in $v_2$ in experimental data. 

\item Event-by-event fluctuations of the initial size, responsible for the transverse-momentum fluctuations, are very similar 
in the wounded  quark model and in the wounded nucleon model amended with binary collisions, with wounded quarks yielding 
somewhat larger values for peripheral collisions. 

\item The analysis of the distribution of particles at high multiplicity in p+Pb and p+p collisions at the LHC energies shows, that the needed overlaid 
distribution, taken in the negative binomial form, receives the same parameters. This is needed for the consistency of the approach.

\item For p+Pb collisions, we find that the size and the eccentricity 
of the fireball is similar as in the {\em compact source} implementation  of the wounded nucleon model
that best describes the data after the hydrodynamic evolution. In that way,
we reduce the uncertainty in the initial conditions in small systems
and obtain a more microscopic motivation for mechanism of the entropy deposition in the initial stage.

\item For p+p collisions we generate initial fireballs which later may be used in hydrodynamic studies. It is expected that the resulting 
triangular flow should be much smaller from the elliptic flow.

\item We have also tested the hypothesis that the effective number of partons in the nucleon is smaller or larger than three. 
However, particle production data at the LHC are fairly well described using three partons, identified with constituent (wounded) quarks. 
 
\end{enumerate}

\begin{acknowledgments}
We thank Andrzej Bia\l{}as, Adam Bzdak, Constantin Loizides and Jurgen Schukraft  for useful discussions.
Research supported by the Polish Ministry of Science and Higher Education, (MNiSW), by the National
Science Centre, Poland grants DEC-2015/17/B/ST2/00101 and DEC-2012/06/A/ST2/00390.
\end{acknowledgments}

\appendix

\section{Quark-quark wounding profile \label{sec:inel}}

\begin{figure}[b]
\begin{center}
\includegraphics[width=0.44 \textwidth]{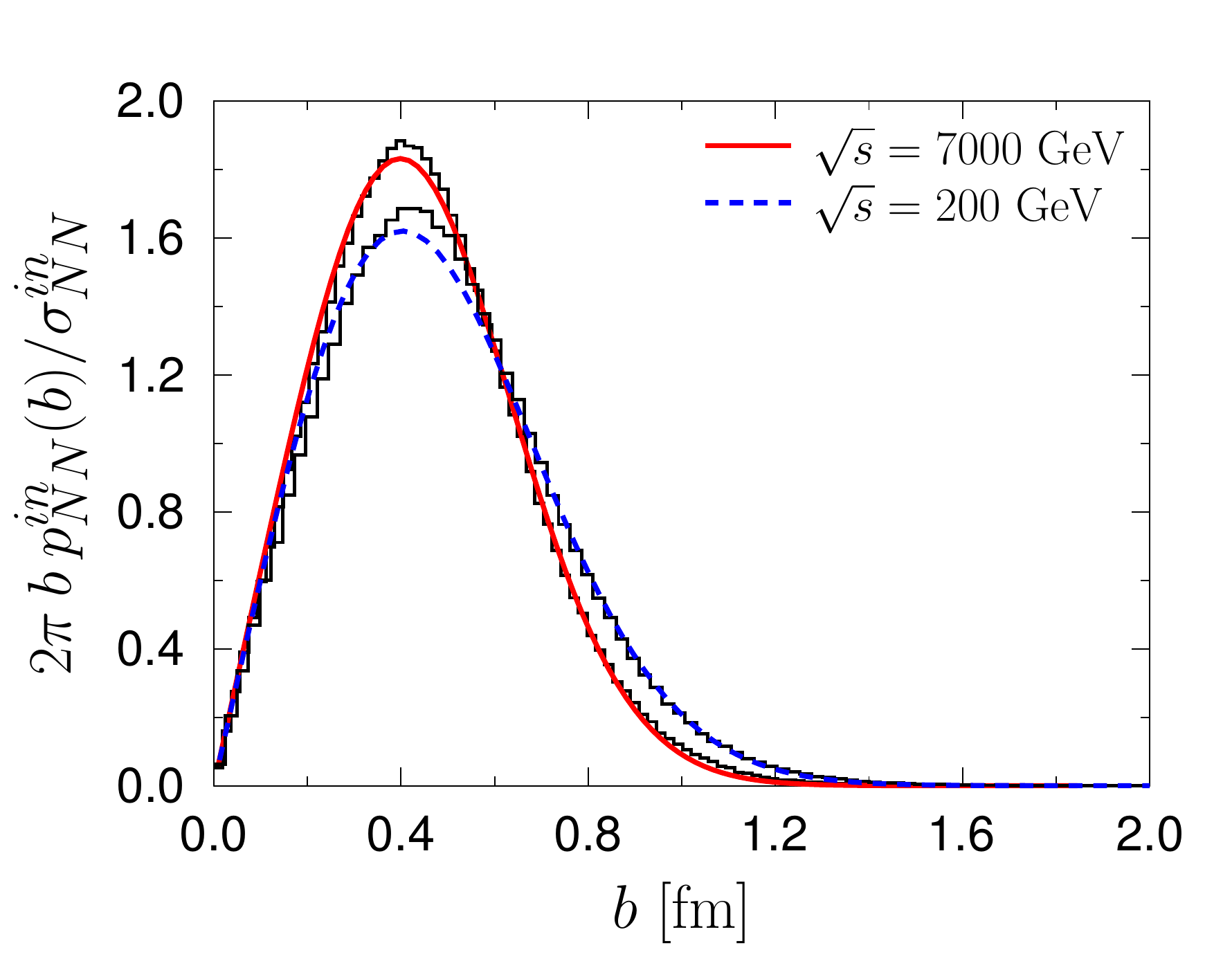} 
\end{center}
\vspace{-5mm}
\caption{Inelasticity profiles $2\pi b p_{NN}^{in}$ obtained from folding three quarks (thick dashed line for $\sqrt{s}=200$~GeV and thick solid line for $\sqrt{s}=7$~TeV), 
compared to the parametrization of Ref.~\cite{Phillips:1974vt} (thin solid lines), plotted 
as functions of the impact parameter $b$.  \label{fig:inel}}
\end{figure}

In this appendix we present in detail our procedure of fixing the inelasticity profile of the quark-quark collisions, based on the 
experimental NN scattering data. The inelasticity profile in NN collisions is defined via the NN scattering amplitude
\begin{eqnarray}
p_{\rm NN}^{\rm in}(b) = 4p ({\rm Im}\, h(b)-p |h(b)|^2),
\end{eqnarray}
where $b$ is the impact parameter, $p$ is the CM momentum of the nucleon, and in the eikonal approximation 
\mbox{$bp = l + 1/2 + {\mathcal O}(s^{-1})$} the scattering amplitude is \mbox{$h(b)=f_l(p)+{\mathcal O}(s^{-1})$}.
Following the lines of Ref.~\cite{Arriola:2016bxa}, we take a working parametrization for $h(b)$ from  Ref.~\cite{Fagundes:2013aja}, whose functional form is based on the 
Barger-Phillips model~\cite{Phillips:1974vt}.

\begin{table}[b]
\caption{Parameters of the quark distribution in the nucleon of Eq.~(\ref{eq:eps}), $r_0$, quark-quark inelastic cross section of Eq.~(\ref{eq:gauss}),  $\sigma_{\rm qq}$, as well as 
parameters $A$ and $\omega$ from Eq.~(\ref{eq:Gamma2}) for various NN collision energies. The last column lists the total inelastic nucleon-nucleon cross section $\sigma_{\rm NN}^{\rm in}$.
\label{tab:inel}}
\medskip
\begin{tabular}{|c|cc|cc|c|}
\hline
$\sqrt{s_{NN}}$ [GeV] & $r_0$ [fm] &  $\sigma_{\rm qq}$ [mb] & $A$ & $\omega$ &$\sigma_{\rm NN}^{\rm in}$ [mb] \\
\hline
17.3  & 0.25 & 5.1  & 0.94 & 0.83 & 31.5 \\
62.4  & 0.27 & 5.8  & 0.92 & 0.79 & 35.6 \\
130   & 0.27 & 6.5  & 0.97 & 0.80 & 38.8 \\
200   & 0.27 & 7.0  & 0.97 & 0.76 & 41.3 \\
2760 & 0.29 & 11.9& 0.99 & 0.57 & 64.1 \\
5020 & 0.30 & 13.6& 1.0   & 0.53 & 70.9 \\
7000 & 0.30 & 14.3& 1.0   & 0.51 & 74.4 \\
\hline
\end{tabular}
\end{table}

When  the nucleon is composed of three constituent quarks, its inelasticity is a folding of the inelasticity of the quark  and the distribution of quarks in the nucleon. 
Here we proceed in an approximate way, using the expressions (\ref{eq:eps},\ref{eq:gauss}). The shape of the resulting NN inelasticity profile is determined be dimensionless ratio 
$\sigma_{\rm qq}/r_0^2$, whereas the normalization, i.e., the NN inelastic cross section \mbox{$\sigma_{NN}^{\rm in} = \int 2\pi b db \, p_{\rm NN}^{\rm in}(b)  $}, 
is controlled with $\sigma_{\rm qq}$. A sample result of the fit for $\sqrt{s_{NN}}$ is shown in Fig.~\ref{fig:inel}.
Richer functional forms of Eqs.~ (\ref{eq:eps},\ref{eq:gauss}) would allow for an even better agreement, but for the present application the accuracy is by far sufficient.

In Table~\ref{tab:inel} we give the  obtained values of model parameters for various collision energies. As the experimental  $p_{\rm NN}^{\rm in}(b)$ may be very well approximated with 
the $\Gamma$ profile of Ref.~\cite{Rybczynski:2013mla},
\begin{eqnarray}
p_{\rm NN}^{\rm in}(b) =  A \Gamma\left (1/\omega, \pi A b^2/(\omega \sigma_{\rm NN}^{\rm in}) \right ) /\Gamma(1/\omega), \label{eq:Gamma2}
\end{eqnarray}
with $\Gamma(a,x)$ denoting the incomplete Euler $\Gamma$ function, we also list parameters $A$ and $\omega$ in Table~\ref{tab:inel}, as well as the inelastic NN cross section 
$\sigma_{\rm NN}^{\rm in}$.

We note that the size parameter $r_0$ in Table~\ref{tab:inel} corresponds to the rms radius of the nucleon (including the CM corrections as obtained from the Monte Carlo 
simulation) equal from 0.70~fm at $\sqrt{s_{NN}}=17.3$~GeV to 0.84~fm at $\sqrt{s_{NN}}=7$~TeV. This range is comparable to 
estimates for the size of the nucleon.

\vfill

\bibliography{hydr}

\end{document}